\documentclass[12pt,letterpaper]{article}

\usepackage[margin=1in]{geometry}
\usepackage[english]{babel}
\usepackage{setspace}
\usepackage{graphicx}
\usepackage{subfigure}
\usepackage{lscape}
\usepackage{fancybox,color}
\usepackage{epstopdf}
\usepackage[capposition=top]{floatrow}
\usepackage{float}
\usepackage{enumitem}
\usepackage{longtable}
\usepackage{tabularx,calc}
\usepackage{multirow}
\usepackage{booktabs}
\usepackage{acro}
\usepackage{mathtools}
\usepackage{rotating}
\usepackage{caption}
\usepackage{amsmath,amsthm,amssymb,amsbsy,bm}
\usepackage[authoryear]{natbib} 
\usepackage[colorlinks=true,linkcolor=blue,citecolor=blue,urlcolor=blue]{hyperref}

\newtheorem{theorem}{Theorem}[section]
\newtheorem{proposition}[theorem]{Proposition}
\newtheorem{lemma}[theorem]{Lemma}
\newtheorem{corollary}[theorem]{Corollary}
\newtheorem{assumption}[theorem]{Assumption}

\newtheorem{assumptionalt}{Assumption}[theorem]
\newenvironment{assumptionp}[1]{
  \renewcommand\theassumptionalt{(#1)}
  \assumptionalt
}{\endassumptionalt}
\let\mc\mathcal
\let\bs\boldsymbol
\let\b\mathbf

\newcommand{\R}{\mathbb{R}}
\newcommand{\N}{\mathbb{N}}
\newcommand{\E}{\mathbb{E}}
\newcommand{\pr}{\mathbb{P}}
\newcommand{\Z}{\mathbb{Z}}	% Integer
\newcommand{\F}{\mathcal{F}}	% Calligraphic F for a sigma algebra
\newcommand{\var}{\mathrm{Var}}

\newcommand{\vect}[1]{\mathrm{vec}\left(#1\right)}

\newcommand{\diag}{\operatorname{diag}}
\newcommand{\trace}{\operatorname{trace}}
\newcommand{\rank}{\operatorname{rank}}
\newcommand{\cip}{\xrightarrow{p}}

\DeclareMathOperator*{\argmin}{arg\,min}

\title{
Efficient estimation of structural VARs with higher-order moments
}
 \author{Sascha  A. Keweloh\thanks{Corresponding author. TU Dortmund, Department of Economics, D-44221 Dortmund, Germany. Sascha Keweloh gratefully acknowledges financial support by TRR 391 “Spatio-temporal Statistics for the Transition of Energy and Transport” (520388526)  by the Deutsche Forschungsgemeinschaft (DFG, German Research Foundation).} 
  \and
 Shu Wang\thanks{University of G\"ottingen, Chair of Econometrics. European University Viadrina Frankfurt, Department of Statistics, Gro{\ss}e Scharrnstra{\ss}e 59, D-15230 Frankfurt (Oder), Germany.  Shu Wang gratefully acknowledges financial support
by the DFG (HE 2188/17-1 \& HE 2188/17-2). }}
\date{\today}

\begin{document}
\maketitle

\begin{abstract}
\begin{singlespace}
     This paper provides a practical standard for efficient implementation of GMM estimation of structural vector autoregressions identified via higher-order moments. We establish conditions under which the structural estimator is adaptive to reduced-form estimation and the conventional weighting is asymptotically optimal. Despite this optimality, its feasible implementation can be unstable because the covariance of the moments involves many higher-order co-moments. We propose exploiting the same identifying assumption for a structured, bias-corrected covariance estimation under both mutual independence and the weaker mean independence, which also allows for common volatility. The resulting estimator is both asymptotically and computationally efficient. Monte Carlo experiments show substantial gains over conventional GMM and performance comparable to likelihood-based estimators under independent shocks. Under common volatility, the mean-independence estimator remains close to the truth while competing estimators are biased. An oil-market application illustrates the practical gains of the approach.
\end{singlespace}
\end{abstract}

\noindent%
	\vfill

	\newpage 
	
\section{Introduction}

Identification by non-Gaussianity has become an important statistical approach to identifying structural vector autoregressive models (SVARs), especially in applications where theory-based restrictions or external instruments are unavailable or controversial. In such settings, higher-order moment conditions of non-Gaussian structural shocks provide identifying information that can be used for identification. This line of work has grown rapidly and now includes a broad class of moment-based and related estimators for SVARs identified from independence restrictions, see, e.g.,  \cite{lanne2021gmm}, \cite{keweloh2020generalized},   \cite{guay2021identification}, \cite{mesters2022non},  \cite{amengual2022moment}, \cite{karamysheva2022we},   \cite{lanne2022statistical},  \cite{lanne2022identifying}, \cite{keweloh2023monetary},   \cite{drautzburg2023refining}, \cite{gourieroux2025generalized}, and \cite{keweloh2026uncertain}.

While the approach of generalized method of moments (GMM) promises a  flexible way to incorporate identifying moment conditions without specifying the underlying density of the shocks, the practical implementation of such an estimator faces two efficiency problems.
First, structural estimation is commonly implemented by plugging into the higher-order moment conditions the estimated VAR residuals, 
which often depend on a large number of reduced-form slope parameters. For example, in a six-dimensional VAR(12) for monthly variables, in order to estimate $36$ structural parameters, one has to estimate $6^2\times12=432$ reduced-form slope parameters. 
It is therefore crucial whether the structural estimator is adaptive to the reduced-form estimation, i.e., whether inference about the structural parameters and efficient weighting using the  inverse moment-covariance (denoted as $S_{gg}^{-1}$) can be carried out as if the reduced-form parameters were perfectly known. 
Second, even if the usual inverse moment-covariance weight remains asymptotically optimal, due to the large number of higher-order moment conditions involved, estimating the moment-covariance $S_{gg}$ itself is difficult.
For example, in the a six-variable SVAR using second- to fourth-order moment conditions, the matrix $S_{gg}$ contains 2,835 co-moments of order four to eight! 
The resulting noise in the estimated weight can therefore undermine the gains promised by asymptotic efficiency. 

This paper addresses the efficiency question by establishing conditions for the optimality of the conventional weighting and subsequently developing a feasible and stable estimator based on a structured, bias-corrected covariance estimation. 
Together, the paper contributes to the literature by providing a standard for practical implementation of GMM estimation of SVARs via higher-order moments.

We account for the two-step nature of SVAR estimation.
The reduced-form AR slopes are estimated by least squares, and the structural moments are then evaluated at the fitted residuals. Estimation of these residuals can affect both the limiting distribution of the structural estimator and its optimal weight, it is therefore not obvious whether the weighting $S_{gg}^{-1}$, which ignores reduced-form estimation, is in fact asymptotically optimal. The issue is recognized in, among others, \cite{gourieroux2020identification, drautzburg2023refining, lanne2022identifying}.
\cite{drautzburg2023refining}.
Our first contribution is to derive the general joint asymptotic distribution of the reduced-form and structural estimators and to characterize the corresponding efficient weighting matrix.
Our analysis yields an affirmative answer for the benchmark two-step SVAR-GMM estimator: the usual weighting $S_{gg}^{-1}$, the inverse covariance of the structural moment functions, can be genuinely optimal in a wide class of models. 
The key is to estimate only the AR slopes in the first step. In a stable, centered VAR without an intercept, the expected Jacobian of the structural moments with respect to these slopes is zero. The structural estimator is therefore adaptive. Thus, even though many slopes may have to be estimated, their uncertainty does not alter the optimal weight for the structural step.

Despite its asymptotic optimality, it remains difficult to estimate $S_{gg}$ accurately in finite samples to realize efficiency gains. The large number of higher-order cross-moments makes direct estimation of an ostensibly optimal weighting matrix unstable in the sample sizes typically encountered in applied work. The difficulty is emphasized in \cite{lewis2025review}, while the computational burden of large moment-covariance inversions is also discussed in \cite{gourieroux2025generalized}.
We show that in Monte Carlo simulations, the weak finite-sample performance of GMM documented in recent studies \citep[see, e.g.,][]{herwartz2023srhl, HW2024jae, hafner2025kml,fiorentini2026single} can be explained by the fact that the weighting matrix is poorly estimated.
Our second contribution is to introduce a structured covariance estimation by exploiting the same assumptions that justify identification to improve estimation of the efficient weighting, through which asymptotic efficiency becomes materialized and thus relevant in practice.
 A key point is that when, for example, mutual independence of the shocks is the maintained identifying assumption, the same assumption should also be used when estimating the efficient weighting. Under such an assumption, the higher-order co-moments entering $S_{gg}$ factorize into products of lower-order moments. As a result, the covariance matrix of the moment conditions can be estimated from a much \textit{smaller} set of \textit{lower-order} marginal moments. In the six-variable example, the corresponding representation requires only 36 moments of order \textit{one to six}, in place of 2,835 co-moments of order \textit{four to eight}. This reduction makes the efficient weighting  feasible and stable in the sample sizes of practical interest.

In applications where the assumption of mutually independent shocks appears too restrictive, GMM can use moment restrictions justified under weaker assumptions such as that the shocks are only independent in their means. The assumption of mean independence is compatible with an empirically relevant setting in which structural shocks may share a common volatility \citep{montiel2022svar, mesters2022non, keweloh2026uncertain}. We propose a structured covariance estimator under the weaker mean independence assumption, which leads to smaller, but still substantial and practically meaningful reduction of number of co-moments. In the six-variable example, the number of co-moments to be estimated can still be reduced by 85\% using the proposed approach.

Additionally, we show that, even with the population optimal weight, the expected criterion locally favors candidate shocks with variances below one in finite samples. Pre-whitening would remove these scale directions from the structural optimization, but it also estimates covariance parameters whose uncertainty generally enters the structural influence function, so that the same adaptivity and optimal-weight argument cannot simply be transferred to that procedure (see Appendix~\ref{app:prewhitening}). We instead address the scaling distortion directly through a continuous scale updating estimator (CSUE). The bias correction is computationally efficient, since we correct only the part of the weighting responsible for the scaling distortion.
This retains the asymptotic efficiency of the estimator and avoids repeatedly estimating and inverting the entire covariance matrix.

The Monte Carlo experiments demonstrate the performance gains of the suggested implementation in finite samples. For the same identifying moment conditions, structured weighting produces a dramatic improvement in the criterion evaluated at the true parameter. In the four-variable design at $T=100$, the GMM criterion falls from approximately 50 under the conventional weighting scheme to 0.56 under the mutual-independence construction, a reduction of almost two orders of magnitude. With structured covariance estimation and continuous scale updating for bias correction, the resulting estimator performs on the same level as Student-$t$ pseudo-maximum likelihood and the flexible discrete location scale mixture of normals (DLSMN) estimator of \citet{fiorentini2023discrete} under independent shocks. When shocks share a persistent common volatility, the mean-independence estimator remains centered close to the true impact responses, while the competing likelihood-based estimators exhibit pronounced bias and produce confidence intervals with severe undercoverage.

The application revisits the oil-market SVAR of \citet{keweloh2026uncertain}. Based on the weaker mean independence assumption, the benchmark estimator detects significant co-movements in the variance of oil supply and oil-specific demand shocks, where the latter shock reflects a precautionary motive of holding oil and predicts a decline in economic activities \citep[see, e.g.,][for similar evidence]{Kilian_Murphy_2014, Anzuini_2015, hafner2025kml}. This is consistent with the argument in the recent study of \cite{HW_2024_HCKML} that precautionary demand for oil can exhibit increased volatility amidst heightened uncertainty surrounding oil supply due to factors such as rising geopolitical tensions, warfare, and sanctions. Compared to the conventional GMM identification that assumes mutual independence, the proposed estimator  yields narrower confidence intervals throughout the reported impulse responses and, in particular, sharpens the evidence regarding stock market responses to unexpected oil supply disruptions.

The remainder is organized as follows. Section~\ref{sec:svar_gmm} introduces the model. Section~\ref{sec:asymptotic-theory} derives the joint asymptotic distribution and the optimal weight. Sections~\ref{sec: estimating s0 and g0} and~\ref{sec: bias} develop structured covariance estimation and continuous scale updating. Section~\ref{sec: Finite sample performance} presents the Monte Carlo evidence, Section~\ref{sec:Application} contains the oil-market application, and Section~\ref{sec:conclusion} concludes. The appendices collect proofs, the analysis of pre-whitening, implementation details, and additional simulation and application results.

% \clearpage
\section{Identification of SVAR via higher-order moments}
\label{sec:svar_gmm}

% This section briefly summarizes  how    higher-order moment conditions can be used to identify and estimate the SVAR. A detailed description can be found in \cite{lanne2021gmm}, \cite{keweloh2020generalized},    \cite{guay2021identification},  or \cite{lanne2022identifying}.

We consider a finite-order structural VAR($p$) model
\begin{align}
\label{eq:SVAR}
	y_t = A_1 y_{t-1} + \ldots + A_p y_{t-p} + u_t,
    \qquad
	u_t = B_0 \varepsilon_t.
\end{align}
where $y_t \in \R^n$ is a vector of endogenous variables, and $\varepsilon_t = (\varepsilon_{1,t},...,\varepsilon_{n,t})'$ contains the structural shocks. The structural impact matrix $B_0$ is non-singular. We make the usual assumption on the data generating process (DGP):
\begin{assumption}{\ }\label{ass:base}
	\begin{enumerate}[label=(\roman*)]
		\item The VAR process in \eqref{eq:SVAR} is causal and stable with $p \in \N$ being fixed and known, s.t., $A(z) = I_n - \sum^p_{i = 1} A_i z^i$ satisfies $\det A(z) \neq 0$ for all $|z|\leq 1$. The process $(y_t)_t$ is initialized from the stationary distribution, and $B_0 \in \mathbb{B} := \{B\in  \R^{n \times n}: \det B \neq 0 \}$. 
		\item The sequence of structural shocks $(\varepsilon_t)_{t\in\Z}$ is iid and satisfies
        $\E[\varepsilon_t] = 0,$  $\E[\varepsilon_t\varepsilon_t']=I_n$.
	\end{enumerate}
\end{assumption}
We work with a causal and stable VAR driven by an iid sequence of structural shocks.  
This setting allows us to characterize the influence functions of the reduced-form and structural estimators, and hence to see clearly the role played by the reduced-form estimation uncertainty for structural identification. As we show next, under this setting the two-step GMM estimation has a nice structure, which yields an adaptivity result for the structural estimator and implies that the efficiency can be obtained by a usual weighting. This same structure is then exploited in Section~\ref{sec: estimating s0 and g0} and~\ref{sec: bias} to develop a scale-stabilized estimator for the efficient weighting and the limiting variance with improved finite-sample properties.

Under Assumption~\ref{ass:base}, $(y_t)_t$ is second-order stationary and has a Wold Moving Average (MA) representation with absolutely summable coefficient matrices:
\begin{align*}
	y_t = \sum^\infty_{h=0} \Phi_h u_{t-h} = \sum^\infty_{h=0} \Theta_h \varepsilon_{t-h}, 
    \qquad
    \Phi(L) = A(L)^{-1}, \; \Theta_h = \Phi_hB_0.
\end{align*}
The structural MA representation $y_t=\Theta(L)\varepsilon_t$ is particularly useful for tracing the dynamic effects of structural shocks on $y_t$ through the impulse response functions (IRFs).

Denote the VAR parameters in \eqref{eq:SVAR} by $\alpha := \text{vec}([  A_1,...,\  A_p]) \in \R^{(n^2p)\times 1}$, define $\b y = \vect{y_1, \ldots, y_T}$, $Z_t = (y_t', \ldots, y_{t-p+1}')' \in \R^{np\times 1}$, and $Z = (Z_0, \ldots, Z_{T-1})$. The reduced-form model can be written as $y_t = (Z_{t-1}' \otimes I_n) \alpha + B_0 \varepsilon_t$.  Suppose $(y_t)_{t=-p+1}^T$ are observations on an $n$-dimensional random vector and the correct lag order $p$ is known. Typically, a structural VAR is estimated in two steps.  First, the reduced-form parameter $\alpha$ is estimated via least squares (LS) by solving the normal equation
\begin{equation}
    \label{eq:m}
\frac1T \sum^T_{t=1}m_t(\alpha)=\bs 0, \quad \text{with } \;   m_t(\alpha) = (Z_{t-1} \otimes I_n)u_t(\alpha), 
\end{equation}
which leads to the LS estimator with the familiar form $\hat \alpha_T = [(ZZ')^{-1}Z\otimes I_n]\b y$. 
Let $u_t(\hat\alpha_T)$ denote the reduced-form residual implied by $\hat\alpha_T$.
In the second step, the simultaneous interaction $u_t = B_0 \varepsilon_t$ is estimated based on the reduced-form residuals $u_t(\hat\alpha_T)$. Denote the structural parameter $\beta = \text{vec}(B)$, $B \in \mathbb{B}$ and the innovations as obtained by %demixing the reduced form residuals
\[
e(\alpha, \beta)_t:= B(\beta)^{-1} u_t(\alpha) =  B(\beta)^{-1} \left(y_t - (Z_{t-1}' \otimes I_n)\alpha \right).
\]
Denote the full parameter vector by $\theta = (\alpha', \beta')'$ and its true value by $\theta_0 = (\alpha_0', \beta_0')'$. 
At $\theta = \theta_0$, the  innovations $e(\theta_0)_t$ coincide with the structural shocks.

\begin{table}[t]
\centering
\caption{Illustration of moment conditions}
\label{Table: Moment conditions }
\small
\renewcommand{\arraystretch}{1.15}

\begin{tabular}{
@{}r@{\hspace{0.5em}$\implies$\hspace{0.5em}}l
@{\hspace{1em}}
r@{\hspace{0.5em}$\implies$\hspace{0.5em}}l@{}
}
\toprule
\multicolumn{2}{c}{Covariance / second-order conditions}
&
\multicolumn{2}{c}{Coskewness / third-order conditions}
\\
\midrule

$\E[\varepsilon_1^2]=1$
& $\E[e_1(\theta)^2]\overset{!}{=}1$
& $\E[\varepsilon_1^2\varepsilon_2]=0$
& $\E[e_1(\theta)^2e_2(\theta)]\overset{!}{=}0$ \\

$\E[\varepsilon_n^2]=1$
& $\E[e_n(\theta)^2]\overset{!}{=}1$
& $\E[\varepsilon_1\varepsilon_2^2]=0$
& $\E[e_1(\theta)e_2(\theta)^2]\overset{!}{=}0$ \\

$\E[\varepsilon_1\varepsilon_2]=0$
& $\E[e_1(\theta)e_2(\theta)]\overset{!}{=}0$
& $\E[\varepsilon_1\varepsilon_2\varepsilon_3]=0$
& $\E[e_1(\theta)e_2(\theta)e_3(\theta)]\overset{!}{=}0$ \\

\midrule
\multicolumn{4}{c}{
Cokurtosis / fourth-order conditions
} \\
\midrule

$\E[\varepsilon_i^3\varepsilon_j]=0$
& $\E[e_i(\theta)^3e_j(\theta)]\overset{!}{=}0$
& $\E[\varepsilon_i^2\varepsilon_j^2]=1$
& $\E[e_i(\theta)^2e_j(\theta)^2]\overset{!}{=}1$ \\

$\E[\varepsilon_i^2\varepsilon_j\varepsilon_k]=0$
& $\E[e_i(\theta)^2e_j(\theta)e_k(\theta)]\overset{!}{=}0$
& $\E[\varepsilon_i\varepsilon_j\varepsilon_k\varepsilon_\ell]=0$
& $\E[e_i(\theta)e_j(\theta)e_k(\theta)e_\ell(\theta)]\overset{!}{=}0$ \\

\bottomrule
\end{tabular}

\par\smallskip
{\footnotesize
\noindent
\textit{Notes:}
$\varepsilon_i=\varepsilon_{i,t}$ and
$e_i(\theta)=e(\theta)_{i,t}$.
Distinct indices denote distinct shocks.
}
\end{table}

% \textcolor{red}{add independence and mean independence assumptions here to motivate moment conditions and later on we use it to motivate S estimation}
Imposing certain structure on the  (in-)dependencies   of the structural shocks allows to derive moment conditions.
For example, if the structural shocks are mutually uncorrelated with unit variance, the demixing matrix $B$ should yield uncorrelated  innovations with unit variance and therefore, satisfy the second-order moment conditions in Table \ref{Table: Moment conditions }. However, there are not sufficiently many second-order moment conditions to identify the  SVAR, which is the well-known identification problem.
Imposing more structure on the higher-order moment (in-)dependencies  of shocks allows for deriving additional higher-order moment conditions.  
A strong form of such a condition is to assume that the  shocks are mutually independent:
\begin{assumptionp}{MI} \label{ass:mutual_indep}
	The components of  $\varepsilon_{t}$ are mutually  independent with at most one component being Gaussian.
\end{assumptionp}
Assumption~\ref{ass:mutual_indep} has been employed in, among others, \cite{gourieroux2017statistical, lanne2017identification, gourieroux2020identification, hafner2025kml}, and implies that the shocks should  satisfy the third- and fourth-order moment conditions in Table \ref{Table: Moment conditions }. These moment conditions would follow from Assumption~\ref{ass:base}(ii) and add no identifying content if the shocks are jointly Gaussian. In general, the moment conditions can be written as
\begin{align}
\label{eq: f}
\E[g_k(y_t, \theta_0)] = 0 \quad \text{with} 
\quad  
g_k(y_t, \alpha, \beta) = 	g_k(e_t(\theta)) :=\prod_{i=1}^{n} e(\theta)_{i,t}^{k_i}- c(k), 
\end{align}
for $k = 1, \ldots, K$, where both the power $k_i$ and the constant $c(k) \in \{0, 1\}$ depends on the exact moment condition outlined in Table \ref{Table: Moment conditions }.

In some settings, mutual independence in~\ref{ass:mutual_indep} can be too restrictive. A  weaker requirement is that each structural shock is unpredictable in mean from the remaining
shocks:\footnote{The notation $\E[\varepsilon_{i,t} \mid  \varepsilon_{-i,t}]$ is a shorthand for the conditional expectation given the sigma-field generated by the set $\varepsilon_{-i,t}$, i.e., $\E[\varepsilon_{i,t} \mid  \boldsymbol{\sigma}(\{\varepsilon_{j,t}: 1\leq j\leq n, j \neq i\})]$. The statements involving conditional expectations are understood as almost surely (a.s.). }
\begin{assumptionp}{WI} \label{ass:mean_indep}
    The components of $\varepsilon_t$ are independent in mean, $\E[\varepsilon_{i,t} \mid  \varepsilon_{-i,t}] = 0$, $\forall i$, with at most one component being Gaussian.
\end{assumptionp}
Mean independence is compatible with the empirically relevant scenario, in which multiple shocks share a common volatility process $ \varepsilon_{i,t} = \sigma_{t} \eta_{i,t}$  where $\sigma_t$ is a common stochastic scalar volatility process independent of $\eta_{it}$. As argued by \cite{montiel2022svar}, the case can be found in many macro models and there is empirical evidence suggesting that volatilities of certain shocks may co-move. 

An  implication of Assumption~\ref{ass:mean_indep} is that any monomial with at least one component entering linearly has zero expectation. Specifically, if there exists $i\in \{1,\ldots,n\}$ such that $k_i=1$, then
$ 
\E\!\left[\prod_{\ell=1}^n \varepsilon_{\ell,t}^{k_\ell}\right]
=\E\!\left[ \left(\prod_{\ell\neq i}\varepsilon_{\ell,t}^{k_\ell}\right)
\E[\varepsilon_{i,t}\mid \varepsilon_{-i,t}]\right] =0$.
In fact, nearly all co-moment conditions discussed above, except for the symmetric cokurtosis conditions $\E[\varepsilon_{i,t}^2\varepsilon_{j,t}^2]=1$ with $i \neq j$ can be derived under the weaker assumption of mutually mean independent shocks. To distinguish this weaker form of independence from mutual independence, write $\mathbf{2}$ that contains the variance and covariance conditions such that $\sum_{i=1}^{n} k_i = 2, \ \forall k \in \mathbf{2}$,
while $\mathbf{3}$ and $\mathbf{4}$ contain coskewness and  cokurtosis conditions so that 
$\sum_{i=1}^{n} k_i = 3, \ \forall k \in \mathbf{3}$ and $\sum_{i=1}^{n} k_i = 4, \ \forall k \in \mathbf{4}$,
and define
$\mathbf{4}^{WI}:=\mathbf{4}\setminus\left\{ k\in\mathbf{4}: k_i=k_j=2 \text{ for some } i<j \text{ and } k_\ell=0 \text{ for all } \ell \notin\{i,j\} \right\}$.
Thus, $\mathbf{4}^{WI}$ contains the asymmetric fourth-order conditions
$\E[\varepsilon_{i,t}^3\varepsilon_{j,t}]=0$,
$\E[\varepsilon_{i,t}^2\varepsilon_{j,t}\varepsilon_{\ell,t}]=0$, and
$\E[\varepsilon_{i,t}\varepsilon_{j,t}\varepsilon_{\ell,t}\varepsilon_{m,t}]=0$,
but excludes the symmetric cokurtosis conditions.

Recently, several studies have made advancements in the identification of non-Gaussian SVAR models under the weaker assumption of  mean independent shocks. Notably, \cite{mesters2022non}, \cite{keweloh2026uncertain}, and \cite{lanne2026identifying} have provided identification results based on moment-based approaches. \cite{keweloh2026uncertain} and \cite{lanne2026identifying} employ third-order moment conditions, which enable the identification of SVAR models when the shocks exhibit sufficient skewness and zero co-skewness, a property derived from the assumption of mean independent shocks. On the other hand,  \cite{mesters2022non} present more general identification results that are not restricted to third-order moment conditions.

% \blue{update/weg:} However, despite these advancements, the simulations in Section \ref{sec: Finite sample performance} reveal that achieving efficient weighting of the higher-order moment conditions and accurate estimation of the asymptotic variances of the estimator can be challenging without relying on the assumption of independent structural shocks.

\section{Asymptotic efficiency of the SVAR-GMM estimator}
\label{sec:asymptotic-theory}

To derive the joint limit of the reduced-form and structural estimators,
write $\dot g_\varepsilon(\varepsilon_t) := \frac{\partial g(e_t(y_t, \theta_0)) }{\partial e_t'} = \frac{\partial g(\varepsilon_t) }{\partial \varepsilon_t'} $ and define
$G_\alpha \coloneqq \E\left[\frac{\partial g(y_t, \alpha_0, \beta_0) }{\partial \alpha'} \right] = -\E[\dot g_\varepsilon(\varepsilon_t) B_0^{-1}(Z_{t-1}' \otimes I_n)]$,
$G_\beta \coloneqq \E\left[\frac{\partial g(y_t, \alpha_0, \beta_0) }{\partial \beta'} \right] = -\E[\dot g_\varepsilon(\varepsilon_t) (\varepsilon_t' \otimes B_0^{-1})]$.
% \begin{align}
%     G_\alpha &\coloneqq \E\left[\frac{\partial g(y_t, \alpha_0, \beta_0) }{\partial \alpha'} \right] = -\E[\dot g_\varepsilon(\varepsilon_t) B_0^{-1}(Z_{t-1}' \otimes I_n)], \\
%     G_\beta &\coloneqq \E\left[\frac{\partial g(y_t, \alpha_0, \beta_0) }{\partial \beta'} \right] = -\E[\dot g_\varepsilon(\varepsilon_t) (\varepsilon_t' \otimes B_0^{-1})]. 
% \end{align}
We make the following assumptions:
\begin{assumption}{\ } \label{ass:beta}
	\begin{enumerate}[label=(\roman*)]
		% \item The components of  $\varepsilon_{t}$ are mutually  independent.
        \item It holds $\E[g(y_t, \theta_0)] = \bs 0$, where $g_t(\theta) \equiv g(y_t, \theta)  = (g_1(y_t, \theta), ..., g_K(y_t, \theta))'$ and $\rank G_\beta = n^2$ with $K \geq n^2$.
        \item There exists a $K \times K$ positive definite matrix $W$ and $\|\hat W_T - W\| \cip 0$.
		\item The map $(\alpha, \beta) \mapsto g(., \alpha, \beta)$ is continuous a.s.\ and continuously differentiable in a neighborhood $\mc N$ of $\theta_0$ and $\E \sup_{(\alpha, \beta) \in \mc N}\left\|\frac{\partial g(y_t, \alpha, \beta)}{\partial \alpha'} \right\|  < \infty$, $\E  \sup_{(\alpha, \beta) \in \mc N}\left\|\frac{\partial g(y_t, \alpha, \beta)}{\partial \beta'} \right\|  < \infty.$
        \item For some $\delta>0$ and $r^*=\max_{1\le k\le K}\sum_{i=1}^n k_i$, $\E\|\varepsilon_t\|^{(2+\delta)r^*}<\infty$
	\end{enumerate}
\end{assumption}
Assumption~\ref{ass:beta} only requires the selected moment conditions to be valid at $\theta_0$ and sufficient ($K \geq n^2$) and informative enough ($\rank G_\beta=n^2$) for identification.\footnote{In particular, identification requires that at most one shock is Gaussian. We refer the readers to \cite{lanne2017identification, gourieroux2017statistical}, and \cite{keweloh2026uncertain} for formal arguments establishing identification under~\ref{ass:mutual_indep}, and under the weaker assumption~\ref{ass:mean_indep}, respectively.}
If at least one cokurtosis condition is imposed, \ref{ass:beta}(iv) requires the existence of at least eighth-order moments.  
It is also worth stressing that the asymptotic properties derived below, including efficiency, adaptivity and optimal weighting, do not require mutual independence~\ref{ass:mutual_indep}. 
Specifically, the statements remain valid under the weaker assumption~\ref{ass:mean_indep} and can thus accommodate the a common volatility process as considered by \cite{montiel2022svar}, as long as the imposed moment conditions involve only monomials with at least one component entering linearly (except for the covariance normalization), i.e., $\mathbf{2}\cup\mathbf{3}\cup\mathbf{4}^{WI}$.\footnote{
The arguments in the proof carry over if Assumption~\ref{ass:base} is respected, which holds true if $(\sigma_t, \eta_t)_t$ is an iid sequence with $\sigma_t$ independent of $\eta_t$ satisfying $\E[\sigma_t]=1$ and the moment conditions in \eqref{ass:beta}(iv).}
For a general weighting matrix $\hat W_T$ satisfying \ref{ass:beta}(ii), the GMM estimator for $\hat\beta_T$ for the structural parameter is 
\begin{align}\label{eq:beta_GMM}
	\hat\beta_T    := \argmin \limits_{\beta} \bar g_T(y_t, \hat\alpha_T, \beta)'\hat W_T \bar g_T(y_t, \hat\alpha_T, \beta), 
\end{align}
where $\bar g_T(y_t, \alpha, \beta)= \frac{1}{T}\sum_{t=1}^{T} g(y_t, \alpha, \beta)$ is the sample moment function and $\hat\alpha_T$ the LS estimator for the reduced-form parameter. We start by presenting the general form of the asymptotic distribution of the SVAR-GMM estimator:
\begin{proposition}\label{thm:asymptotics}
Under Assumptions \ref{ass:base} and~\ref{ass:beta},
\begin{align*}
\sqrt T
\binom{\hat\alpha_T-\alpha_0}{\hat\beta_T-\beta_0}
\ &\Rightarrow \
\mc N(0,V_\theta), &
\qquad
V_\theta &=
\begin{pmatrix}
V_\theta^{(\alpha,\alpha)} &
V_\theta^{(\alpha,\beta)} \\
V_\theta^{(\beta,\alpha)} &
V_\theta^{(\beta,\beta)}
\end{pmatrix},\\
    V_\theta^{(\alpha,\alpha)}
    &= \Gamma^{-1} \otimes \Sigma_u,  & 
    V_\theta^{(\alpha,\beta)}
    &= - \left[(\Gamma^{-1} \otimes \Sigma_u ) G_\alpha' + M^{-1}S_{mg} \right] W G_\beta (G_\beta'WG_\beta)^{-1}, \\
    V_\theta^{(\beta,\alpha)}
    &= \left(V_\theta^{(\alpha,\beta)} \right)', & 
    V_\theta^{(\beta,\beta)}
    &= (G_\beta'WG_\beta)^{-1}G_\beta'WS_\perp WG_\beta(G_\beta'WG_\beta)^{-1},
\end{align*}
with $\Gamma := \E[Z_{t-1} Z_{t-1}']$, $\Sigma_u := \E[u_t u_t'] = B_0 B_0'$,
$S_{gg}:=\E[g_t(\alpha_0, \beta_0) g_t(\alpha_0, \beta_0)']$, 
$S_\perp := \E\left[g_t^{\perp}(\alpha_0, \beta_0) g_t^{\perp}(\alpha_0, \beta_0)' \right] = S_{gg} 
+ G_\alpha M^{-1} S_{mg}
+ S_{mg}' M^{-1} G_\alpha'
+ G_\alpha (\Gamma^{-1} \otimes \Sigma_u ) G_\alpha'$ and $g_t^{\perp}(\alpha_0, \beta_0) := g_t(\alpha_0, \beta_0) + G_\alpha M^{-1} m_t(\alpha_0).$
\end{proposition}

As shown in the proof of Proposition~\ref{thm:asymptotics}, the influence function of the GMM estimator for the structural parameter is $g_t^{\perp}(\alpha_0, \beta_0) $, rather than $g_t(\alpha_0, \beta_0)$. 
% To see this, let $M:= \Gamma \otimes I_n$, note that
% \begin{align*}
%     \sqrt{T}(\hat\beta_T-\beta_0)
% &=
% -(G_\beta'WG_\beta)^{-1}G_\beta'W
% \left(
% \frac1{\sqrt T}\sum_{t=1}^T g_t(\alpha_0, \beta_0)
% +
% G_\alpha \sqrt{T}(\hat\alpha_T-\alpha_0)
% \right)
% +o_p(1)\\
% &= -(G_\beta'WG_\beta)^{-1}G_\beta'W \frac1{\sqrt T}\sum_{t=1}^T \left(g_t(\alpha_0, \beta_0) + G_\alpha M^{-1} m_t(\alpha_0) \right) + o_p(1)
% \end{align*}
% where $ M^{-1} m_t(\alpha_0) $ is the influence function of the LS estimator for $\alpha$. 
As a consequence, the optimal weighting leading to asymptotically efficient estimation is the inverse of the variance of the \textit{orthogonalized} moment functions $S_\perp^{-1}$ with $S_\perp = \E[g_t^{\perp}(\alpha_0, \beta_0) g_t^{\perp}(\alpha_0, \beta_0)']$. If the process $(y_t)_t$ has zero mean and the VAR model contains no intercept, as displayed in Eq.~\eqref{eq:SVAR}, then
$\E[Z_{t-1}] = \left(\sum_{j=0}^\infty \left(\Theta_{j} \E[\varepsilon_{t-j-1}]\right)', \ldots, \sum_{j=0}^\infty \left(\Theta_{j}  \E[\varepsilon_{t-j-p}]\right)' \right)' = \bs 0_{np \times 1}.$ In this case, Lemma~\ref{lem:CLT_iid} shows that the moment functions for estimating the reduced-form and structural parameters are mutually orthogonal, i.e., $S_{mg} = \bs 0_{n^2p \times K}$, so that $S_\perp = S_{gg}  +G_\alpha (\Gamma^{-1} \otimes \Sigma_u ) G_\alpha'$, which is larger than $S_{gg}$ in a positive semi-definite sense. Hence, the optimal weighting is given by 
$W^*:=S_{\perp}^{-1}$ and the limiting variance of the  corresponding estimator is $(G_\beta' S_{\perp}^{-1} G_\beta)^{-1}$. 
% This is summarized in the following Corollary:
% \begin{corollary}\label{corol:S_mg}
% Under Assumptions~\ref{ass:base} and~\ref{ass:beta}, it holds that $S_{mg} = \bs 0_{n^2p \times K}$.
% Consequently,
% \[
% V_\theta
% =
% \begin{pmatrix}
% \Gamma^{-1} \otimes \Sigma_u &  -(\Gamma^{-1} \otimes \Sigma_u ) G_\alpha' WG_\beta(G_\beta'WG_\beta)^{-1}\\
% -(G_\beta'WG_\beta)^{-1}G_\beta' WG_\alpha (\Gamma^{-1} \otimes \Sigma_u ) & (G_\beta'WG_\beta)^{-1}G_\beta'WS_\perp WG_\beta(G_\beta'WG_\beta)^{-1}
% \end{pmatrix},
% \]
% In this case, the optimal weighting matrix is $W^*:=S_{\perp}^{-1} = \left( S_{gg}  +G_\alpha (\Gamma^{-1} \otimes \Sigma_u ) G_\alpha' \right)^{-1} $ and the corresponding efficient GMM estimator $\hat\beta_T^*$ satisfies $\sqrt T (\hat\beta_T^*-\beta_0)
% \ \Rightarrow \
% \mc N(0, (G_\beta' S_{\perp}^{-1} G_\beta)^{-1})$.
% \end{corollary}

Notably, $G_\alpha$ determines whether the estimation of the reduced-form parameter has an impact on the estimation of the structural parameter. Specifically, the influence function for $\hat \alpha_T$ appears in the influence function for $\hat \beta_T$ if the Jacobian $G_\alpha \neq \bs 0$.  By serial independence of $\varepsilon_t$ under~\ref{ass:base}(ii), any measurable function of $\varepsilon_t$  is independent of $\F_{t-h} = \sigma(\varepsilon_{s}: s \leq t-h)$ for all $h \geq 1$, and since $Z_{t-1}$ is $\F_{t-1}$-measurable, we have
\[
G_\alpha
=
-\E[\dot g_\varepsilon(\varepsilon_t)B_0^{-1}(Z_{t-1}'\otimes I_n)]
=
-\E[\dot g_\varepsilon(\varepsilon_t)]\,B_0^{-1}\,(\E[Z_{t-1}']\otimes I_n)
=
\bs 0_{K\times n^2p},
\]
where the last equality follows from
$
\E[Z_{t-1}]  = \bs 0_{np \times 1}.
$
This presents a nice structure that allows us to derive the following corollary:
\begin{corollary}\label{corol:G_a}
Under Assumptions~\ref{ass:base} and~\ref{ass:beta}, 
$g_t^\perp(\alpha_0,\beta_0)=g_t(\alpha_0,\beta_0), \ S_\perp=S_{gg}$, and hence
\[
V_\theta
=
\begin{pmatrix}
\Gamma^{-1}\otimes\Sigma_u & \bs 0_{n^2p\times n^2}\\
\bs 0_{n^2\times n^2p} & (G_\beta'WG_\beta)^{-1}G_\beta'WS_{gg}WG_\beta(G_\beta'WG_\beta)^{-1}
\end{pmatrix}.
\]
%In other words, $\hat\beta_T$ is adaptive such that the $\alpha$ can be treated as if its true value $\alpha_0$ were known.  
The optimal weighting is $W^*:=S_{gg}^{-1}$ and the corresponding efficient estimator satisfies
\[
\sqrt T
\binom{\hat\alpha_T-\alpha_0}{\hat\beta_T^*-\beta_0}
\ \Rightarrow \
\mc N(0,V_\theta^*),
\qquad
V_\theta^*
=
\begin{pmatrix}
\Gamma^{-1}\otimes\Sigma_u & \bs 0_{n^2p\times n^2}\\
\bs 0_{n^2\times n^2p} & (G_\beta'S_{gg}^{-1}G_\beta)^{-1}
\end{pmatrix}.
\]
\end{corollary}
As a consequence, $\hat \beta_T$ has the same limiting behavior, as if the true value $\alpha_0$ were known and the reduced-form residual $u_t$ were observed by the econometrician. Moreover,  the usual optimal weighting $S_{gg}^{-1}$, which is derived by ignoring the estimation uncertainty of the reduced form, becomes genuinely optimal. 
Given the block diagonal structure of $V_\theta$, we obtain the asymptotic distribution of structural IRFs:
\begin{corollary}
    Under the Assumptions of Proposition \ref{thm:asymptotics}, it holds for $h\geq 0$,
    \[
    \sqrt{T} \operatorname{vec}\left(\hat \Theta_h - \Theta_h \right) \ \Rightarrow \ \mathcal{N}(0,V_{\Theta_h}),
    \]
    where $V_{\Theta_h} = C_{h,\alpha} \left(\Gamma^{-1}\otimes\Sigma_u \right)  C_{h,\alpha}' +  C_{h,\beta}   (G_\beta'WG_\beta)^{-1}G_\beta'WS_{gg}WG_\beta(G_\beta'WG_\beta)^{-1}  C_{h,\beta}',$
    with
    \begin{align*}
       C_{h,\alpha} =
       \begin{cases}
           0, &  h=0 \\
           (B_0' \otimes I_n) \sum_{m=0}^{h-1} J (\mathbf{A}')^{h-1-m} \otimes \Phi_m, & h\geq 1
       \end{cases}
       \qquad
       C_{h,\beta} = \frac{\partial \operatorname{vec} \Theta_h}{\partial \beta'} = I_n \otimes \Phi_h,
    \end{align*}
    where $J=(I_n, 0, \dots, 0)$ and $\mathbf{A}$ is the companion matrix of VAR(p) process in \eqref{eq:SVAR}. Furthermore, under the optimal weighting $W^*=S_{gg}^{-1}$, we have
    \[
       V_{\Theta_h} = C_{h,\alpha} \left(\Gamma^{-1}\otimes\Sigma_u \right)  C_{h,\alpha}' +  C_{h,\beta}   (G_\beta'S_{gg}^{-1}G_\beta)^{-1}  C_{h,\beta}' 
    \]
\end{corollary}
Hence, the asymptotic uncertainty of structural IRFs can be  decomposed into two parts, i.e., the uncertainty coming from estimating the reduced form, and that from estimating the structural parameters. An important reason for this  adaptivity result is that the first-step estimator concerns only the reduced-form slope parameter $\alpha$, but not the reduced-form covariance matrix $\Sigma_u := \E[u_t u_t'] = B_0 B_0'$. Instead, we may also estimate the reduced-form scale parameters, use them to pre-whiten the residuals, and then estimate a structural orthogonal matrix from higher-order moments based on the pre-whitened residuals. This is a procedure often implemented in the Independent Component Analysis (ICA) estimation strategies \citep[see, e.g.,][]{gourieroux2017statistical, HW2024jae, hafner2025kml}. However, in that case, the Jacobian with respect to the first-step scale parameter is generically nonzero, so that  the influence function of the structural estimator contains the first-step covariance moments. Consequently, the block-diagonal limit in Corollary~\ref{corol:G_a} is lost, and the usual weighting matrix $S_{gg}^{-1}$ is no longer asymptotically optimal. Appendix~\ref{app:prewhitening} provides a formal discussion and  illustration.

\section{Structured covariance estimation}
\label{sec: estimating s0 and g0}

While the theory developed in the previous section establishes the asymptotic optimality of the weighting $S_{gg}^{-1}$, estimating it (and the expected Jacobian  $G_\beta$) can be quite challenging in practice. Since the moment functions contain higher-order monomials, the covariance matrix $S_{gg}$ involves cross-moments of even higher powers, which are very difficult to estimate reliably in sample sizes relevant for many macroeconomic applications. As we will illustrate in the Monte Carlo section, the underperformance of the GMM estimator relative to other independence-based procedures documented in previous simulation studies \citep[e.g.,][]{hafner2025kml} is driven by the fact that the matrix $S_{gg}^{-1}$, despite being the asymptotically optimal weight, is poorly estimated. In the following, we propose a structured way of estimating the moment covariance by exploiting implications of the underlying identifying assumptions. In Appendix~\ref{app:estimate_jacobian}, we apply the same idea to improve the estimation of $G_\beta$.\footnote{In a GMM setup unrelated to SVAR models, \cite{burnside1996small} propose to impose restrictions of the underlying economic model   on the estimator of $S_{gg}$ and $G_\beta$. See also \cite{amengual2022moment} for a similar idea for deriving a test for independent components.}

In general, the long-run covariance  of two  arbitrary moment conditions  
$ g_k(y_t, \theta) = \prod_{i=1}^{n} e(\theta)_{i,t}^{k_i} - c $
and
$ g_{\tilde{k}}(y_t, \theta) = \prod_{i=1}^{n} e(\theta)_{i,t}^{\tilde{k}_i} - \tilde{c} $ 
with $k,\tilde{k} \in  \mathbf{2} \cup \mathbf{3} \cup \mathbf{4}$ 
, $c=c(k)$, and $ \tilde{c}=c(\tilde{k})  $
at $\theta=\theta_0$ is 
\begin{align} \label{eq: S}
S_{g_k,g_{\tilde{k}}} =& \underset{T \rightarrow \infty }{\lim} 
\E \!\left[T 
\left(\frac{1}{T}  \sum_{t=1}^{T}  g_k(\theta_0) \right)\!
\left( \frac{1}{T} \sum_{t=1}^{T}  g_{\tilde{k}}(\theta_0)  \right) 
\right]\\  \nonumber
=& 
\E\! \left[\prod_{i=1}^{n}  \varepsilon_{i,t}^{k_i+\tilde{k}_i}    \right] 
-  c  \E\! \left[\prod_{i=1}^{n} \varepsilon_{i,t }^{\tilde{k}_i}  \right] 
-  \tilde{c}   \E\! \left[\prod_{i=1}^{n}     \varepsilon_{i,t }^{k_i}  \right] 
+  c \tilde{c} 
\\ \nonumber
+ 
&\sum_{j=1}^{\infty}  
\E\! \left[\prod_{i=1}^{n}  \varepsilon_{i,t}^{k_i}  \varepsilon_{i,t-j}^{\tilde{k}_i}   \right] 
-  c  \E\! \left[\prod_{i=1}^{n} \varepsilon_{i,t-j}^{\tilde{k}_i} \right] 
-  \tilde{c}   \E \! \left[\prod_{i=1}^{n}     \varepsilon_{i,t }^{k_i}  \right] 
+  c \tilde{c}
\\ \nonumber
+ 
&\sum_{j=1}^{\infty}  
\E\! \left[\prod_{i=1}^{n}  \varepsilon_{i,t-j}^{k_i} \varepsilon_{i,t}^{\tilde{k}_i} \right]  
-  c  \E\! \left[\prod_{i=1}^{n}   \varepsilon_{i,t}^{\tilde{k}_i}  \right]   
-  \tilde{c}   \E\! \left[\prod_{i=1}^{n}   \varepsilon_{i,t-j}^{k_i}  \right]   
+  c \tilde{c} , 
\end{align}
where the last equality follows from   identically distributed shocks and  $e(\theta_0)_t=\varepsilon_t$. 
Therefore, with fourth-order moments, i.e., $k,\tilde{k}\in \mathbf{4}$ such that $\sum_{i=1}^{n} k_i = \sum_{i=1}^{n} \tilde{k}_i = 4$, the long-run covariance matrix $S_{g_k,g_{\tilde{k}}}$ contains co-moments of the structural shocks up to order eight. In practice, $S_{g_k,g_{\tilde{k}}}$  in (\ref{eq: S}) can be estimated by replacing $\varepsilon_t$ with $e(\theta)_t$ and some initial estimate  $\theta$  and a HAC estimator \citep{newey1994automatic}.
However, under serial independence in Assumption \ref{ass:base}(ii)  the expression simplifies to 
\begin{align}
\label{eq: S SI}
S_{g_k,g_{\tilde{k}}}^{SI} = \E \! \left[\prod_{i=1}^{n}  \varepsilon_{i,t}^{k_i+\tilde{k}_i}    \right] 
-  c  \E\! \left[\prod_{i=1}^{n} \varepsilon_{i,t }^{\tilde{k}_i}  \right] 
-  \tilde{c}   \E\! \left[\prod_{i=1}^{n}     \varepsilon_{i,t }^{k_i}  \right] 
+  c \tilde{c}.
\end{align}
Thus, under serial independence, the optimal weighting can be computed with the  frequently used estimator $\hat S_{gg,T} =  \frac{1}{T}\sum_{t=1}^{T} g_t g_t'$ for the moment covariance 	$S^{SI}$. Despite such simplification, the covariance  matrix $S_{g_k,g_{\tilde{k}}}^{SI}$ of two   fourth-order moments    is still   of order eight and remains difficult to estimate. For instance, consider the two moment conditions $\E[\varepsilon_{1,t}^3 \varepsilon_{2,t}]=0$ and $\E[\varepsilon_{3,t}^3 \varepsilon_{4,t}]=0$, such that the covariance of both moment conditions is equal to $ \E[\varepsilon_{1,t}^3 \varepsilon_{2,t}\varepsilon_{3,t}^3 \varepsilon_{4,t}] $, a co-moment of order eight.

% If we are willing to impose a stronger  condition than \ref{ass:beta}(i), we can further simplify the estimation of $S_{gg}$:
% \begin{assumption}{\ } \label{ass:mutual_indep}
% 	The components of  $\varepsilon_{t}$ are mutually  independent. 
% \end{assumption}

In many ICA-type estimators, which the GMM procedure is often compared to, a stronger mutual independence Assumption~\ref{ass:mutual_indep} is imposed. If we are willing to accept such an assumption, we can further simplify the estimation of $S_{gg}$ as
\begin{align} 
\label{eq: S SIMI}
S_{g_k,g_{\tilde{k}}}^{SMI} =\prod_{i=1}^{n}  \E \! \left[\varepsilon_{i,t}^{k_i+\tilde{k}_i}    \right] 
-  c   \prod_{i=1}^{n} \E\! \left[ \varepsilon_{i,t }^{\tilde{k}_i}  \right] 
-  \tilde{c}   \prod_{i=1}^{n}  \E\! \left[   \varepsilon_{i,t }^{k_i}  \right] 
+  c \tilde{c} ,
\end{align}
Therefore, under serially and mutually independence, higher-order co-moments can be transformed into a product of lower-order moments, i.e., 
$\E[\varepsilon_{1,t}^3 \varepsilon_{2,t} \varepsilon_{3,t}^3 \varepsilon_{4,t}]$ of two moment conditions, 
$\E[\varepsilon_{1,t}^3 \varepsilon_{2,t}]=0$ and $\E[\varepsilon_{3,t}^3 \varepsilon_{4,t}]=0$, 
simplifies to  $ \E[\varepsilon_{1,t}^3] \E[\varepsilon_{2,t}] \E[\varepsilon_{3,t}^3] \E[\varepsilon_{4,t}] $, which is a product of moments of order one and three. 
In general, the covariance matrix under serially independent shocks $S^{SI}$ requires estimating co-moments of $\varepsilon_t$ of order four to eight, while the covariance matrix under serially and mutually independent shocks $S^{SMI}$ requires estimating moments of $\varepsilon_t$ of order one to six. 
Table \ref{Table: moments of S} shows the number of co-moments of $\varepsilon_t$ contained in $S^{SI}$ and $S^{SMI}$ for an estimator using all second- to fourth-order moment conditions. The number of co-moments increases quickly with the dimension of $\varepsilon_t$. For instance, in an SVAR with two variables  $S^{SI}$ requires   estimating five co-moments of order eight, but in six-dimensional system, this number grows to $1251$ co-moments of order eight.  In contrast, the number of moments in $S^{SMI}$ grows linearly. Thus, leveraging independence for covariance estimation is especially meaningful in larger systems.

Nevertheless, in scenarios, where mutual independence is inadequate due to concerns like common volatility, the weaker form of mean independence in Assumption~\ref{ass:mean_indep} can also be exploited to obtain a partial simplification of $S_{gg}$. 
Similar to Equation \eqref{eq: S SIMI},
for $k,\tilde{k}\in \mathbf{2}\cup\mathbf{3}\cup\mathbf{4}^{WI}$, components in $S_{gg}$ under the weaker independence Assumption~\ref{ass:mean_indep} are
\begin{align}
\label{eq: S SWI}
S_{g_k,g_{\tilde{k}}}^{SWI}
&=
\begin{cases}
0, & \text{if } k_i+\tilde k_i=1 \text{ for at least one } i,\\[4pt]
\E\!\left[\prod_{i=1}^n\varepsilon_{i,t}^{k_i+\tilde k_i}\right]-c\tilde c, & \text{otherwise}.
\end{cases}
\end{align}
Unlike simplification under \ref{ass:mutual_indep}, where all higher-order co-moments factorize into products of lower-order marginal moments, mean independence imposes zero restrictions on co-moments, whose exponent vector contains at least one entry equal to one.\footnote{In a small sample, the estimated $S_{g_k,g_{\tilde{k}}}^{SWI}$ may not lead to a positive definite matrix. A common algorithm for restoring positive definiteness can be applied, see Appendix \ref{sec: appendix mi regularization} for details.} For example, the covariance between the two asymmetric fourth-order moment conditions $\E[\varepsilon_{1,t}^3\varepsilon_{2,t}]=0$ and $\E[\varepsilon_{3,t}^3\varepsilon_{4,t}]=0$ is $\E[\varepsilon_{1,t}^3\varepsilon_{2,t}\varepsilon_{3,t}^3\varepsilon_{4,t}],$ which is an eighth-order co-moment under $S^{SI}$ but is equal to zero under Assumption~\ref{ass:mean_indep}. %Similarly, $\E[\varepsilon_{1,t}^3\varepsilon_{2,t}\varepsilon_{1,t}\varepsilon_{2,t}\varepsilon_{3,t}\varepsilon_{4,t}] = \E[\varepsilon_{1,t}^4\varepsilon_{2,t}^2\varepsilon_{3,t}\varepsilon_{4,t}]$ is also zero. 
Table~\ref{Table: moments of S SWI} illustrates the reduction obtained by exploiting Assumption~\ref{ass:mean_indep}. For instance, when $n=6$, the number of distinct co-moments entering $S^{SI}$ is $2805$, including $1221$ eighth-order co-moments. Under mean independence,  the corresponding matrix $S^{SWI}$ requires estimating only $416$ co-moments, a reduction of about $85\%$, including $180$ eighth-order co-moments. %Hence, mean independence delivers a substantial reduction in the number of nuisance co-moments that must be estimated. This reduction is necessarily smaller than the one obtained under mutual independence. Under mutual independence, the remaining co-moments factorize into products of lower-order marginal moments, so that the number of objects to estimate grows only linearly with $n$. By contrast, under Assumption~\ref{ass:mean_indep}, even-power co-moments such as $\E[\varepsilon_{1,t}^4\varepsilon_{2,t}^4]$ and $\E[\varepsilon_{1,t}^2\varepsilon_{2,t}^2\varepsilon_{3,t}^2\varepsilon_{4,t}^2]$ are left unrestricted and must be estimated directly.  This is the price paid for relying on a weaker independence restriction. The advantage is that $S^{SWI}$ remains compatible with common stochastic volatility, that would violate mutual independence and hence invalidate the factorized estimator $S^{SMI}$.

\begin{table}[t]
\centering
\caption{Number of moment conditions and co-moments
under mutual and mutual mean independence}
\label{Table: moments of S}
\label{Table: moments of S SWI}

\begin{tabular}{@{}lrrrrr@{}}
\toprule
 & $n=2$ & $n=3$ & $n=4$ & $n=5$ & $n=6$ \\
\midrule

\multicolumn{6}{@{}l}{
\textit{GMM moment conditions}
} \\
Second-order & 3 & 6 & 10 & 15 & 21 \\
Third-order & 2 & 7 & 16 & 30 & 50 \\
Fourth-order (MI/WI)
 & 3/2 & 12/9 & 31/25 & 65/55 & 120/105 \\
$S_{gg}$ dimension (MI/WI)
 & 8/7 & 25/22 & 57/51 & 110/100 & 191/176 \\

\midrule
\multicolumn{6}{@{}l}{
\textit{Distinct co-moments in } $S^{SI}$
} \\
Fourth-order & 5 & 15 & 35 & 70 & 126 \\
Fifth-order & 4 & 18 & 52 & 121 & 246 \\
Sixth-order & 5 & 25 & 80 & 205 & 456 \\
Seventh-order & 4 & 27 & 104 & 305 & 756 \\
Eighth-order (MI/WI)
 & 5/3 & 36/30 & 149/137 & 470/450 & 1251/1221 \\
Total (MI/WI)
 & \textbf{23/21} & \textbf{121/115}
 & \textbf{420/408} & \textbf{1171/1151}
 & \textbf{2835/2805} \\

\midrule
\multicolumn{6}{@{}l}{
\textit{Distinct (co-) moments in $S^{SMI}$/$S^{SWI}$}
} \\
Orders 1--3 (each)
 & 2/-- & 3/-- & 4/-- & 5/-- & 6/-- \\
Fourth-order
 & 2/3 & 3/6 & 4/10 & 5/15 & 6/21 \\
Fifth-order
 & 2/2 & 3/6 & 4/12 & 5/20 & 6/30 \\
Sixth-order
 & 2/3 & 3/10 & 4/22 & 5/40 & 6/65 \\
Seventh-order
 & --/4 & --/15 & --/36 & --/70 & --/120 \\
Eighth-order
 & --/3 & --/15 & --/43 & --/95 & --/180 \\
Total
 & \textbf{12/15} & \textbf{18/52}
 & \textbf{24/123} & \textbf{30/240}
 & \textbf{36/416} \\
\bottomrule

\end{tabular}

\medskip
\begin{minipage}{\linewidth}
\footnotesize
\textit{Notes:}
MI denotes mutual independence; WI denotes
componentwise conditional-mean independence. 
\end{minipage}
\end{table}

%An alternative approach to reduce the number of co-moments in the covariance matrix $S_{gg}$ is to reduce the number of moments of the GMM estimator as proposed in \cite{lanne2021gmm} and \cite{lanne2022statistical}, although omitting moment conditions may lead to a loss of efficiency. \cite{lanne2021gmm} suggest a strategy involving the sequential application of an information criterion to select an appropriate set of moment conditions. Nevertheless, the method may appear impractical for large SVAR models, where the number of potential combinations of moment conditions escalates  rapidly with the dimension of the SVAR. To circumvent the need to exhaustively explore all possible combinations of moment conditions, one could employ theoretical findings to identify redundant higher-order moment conditions. Since such theoretical results are not available in a general setting, the remainder of this paper focuses on efficient estimators using all second- to fourth-order moment.\footnote{ An exception  is \cite{keweloh2025higher}, which establishes that in the special case of a recursive SVAR with independent shocks, all non-bivariate coskewness and cokurtosis conditions, meaning moment conditions involving more than two shocks like for instance $	\E[\varepsilon_{1,t} \varepsilon_{2,t}\varepsilon_{3,t} ]  =0$, are redundant. }

\section{Small-sample bias and continuous scale updating}
\label{sec: bias} 
This section shows that the asymptotically efficient estimator is biased towards innovations with a variance smaller than the imposed unit variance normalization.  To address this, we develop a continuous scale updating estimator (CSUE), which corrects the bias by incorporating a continuously updated scaling term into the weighting matrix. %The CSUE is computationally cheap since it does not require to continuously re-estimating and updating the matrix $S_{gg}$. The resulting estimator is free from bias and remains asymptotically efficient.

\cite{han2006gmm} and \cite{newey2009generalized} show that  for iid observations and a nonrandom weighting matrix $W$  the    expected value of a GMM objective function is 
\begin{align}
\label{eq: avg finite sample loss 1}
\E\!\left[ \bar g_T(y_t, \theta)'W \bar g_T(y_t, \theta) \right] 
=&  \E\!\left[ \sum_{t \neq \tilde{t}}^{ } g_t(\theta)' W g_{\tilde t}(\theta) \right]/ T^2
+ \E\!\left[ \sum_{t }^{ } g_t(\theta)' W g_t(\theta) \right] / T^2
\\
\label{eq: avg finite sample loss 2}
=& (1-T^{-1}) \E [ g_t(\theta)]' W \E[ g_t(\theta)]  + \trace(W S_{gg}(\theta))/T,
\end{align} 
with $ S_{gg}(\theta):= \E[ g_t(\theta)g_t(\theta)']$.
Equation (\ref{eq: avg finite sample loss 2}) decomposes the expected value into a signal  and a noise term, with the former being minimized at the true parameter value $\theta_0$. The noise term, however, is not minimized at $\theta_0$ and in small samples, the noise term can dominate the signal term and lead to a  bias, especially in    models with many moment conditions.

%------------------------------------------------------------------------------------------------%
% Adapted
%------------------------------------------------------------------------------------------------%

% \clearpage

To illustrate the small-sample bias, we keep the reduced-form parameter fixed at its true value
$\alpha_0$ throughout this section and write $\theta = (\alpha_0',\beta')'$ with $\theta_0 = (\alpha_0',\beta_0')'.$ 
For any diagonal matrix with positive entries $D = \diag(d_1,\ldots,d_n),$ let $\beta_D$ denote the structural parameter such that
$B(\beta_D) = B(\beta)D,$ and define the scaled parameter $\theta_D := (\alpha_0',\beta_D')'.$ 
Since $e_t(\theta_D)=D^{-1}e_t(\theta),$ the $k$-th moment condition transforms according to the exponent vector attached to the $k$-th moment condition in \eqref{eq: f}. Note that $B(\beta_{0,D}) = B_0 D$ and for $i =1, \ldots, n$, we have $\var \left(e(\theta_{0,D})_{i,t} \right) = d_i^{-2}$. 

The following proposition shows that, at $B_0$ with the correctly normalized solution, deviating from the normalizing unit variance moment conditions towards innovations with a variance smaller than one reduces the noise term of the asymptotically efficient estimator. Hence the asymptotically efficient SVAR-GMM estimator has small-sample bias toward solutions that correspond to innovations with variance smaller than one.
\begin{proposition}
\label{prop:bias}
Suppose Assumptions~\ref{ass:base} and~\ref{ass:beta} hold, and 
for any positive diagonal matrix $D=\diag(d_1,\ldots,d_n)$, define
$\delta_k(D) := \prod_{i=1}^n d_i^{-k_i},$ 
$\Delta(D) := \diag \left(\delta_1(D),\ldots,\delta_K(D)\right),$
and $\bs c := \left(c(1),\ldots,c(K)\right)'$, then the noise term can be expressed as
\begin{align}
\nonumber
\frac{1}{T}\trace\bigl(W S_{gg}(\theta_D)\bigr)
&=
\frac{1}{T} \sum_{k=1}^K \sum_{j=1}^K \delta_k \delta_j W_{kj} S_{gg, kj}(\theta) 
+
\frac{2}{T}\,
\sum_{k=1}^K \sum_{j=1}^K c(k) (\delta_k - 1) \delta_j W_{kj} \E[g_j(y_t, \theta)] 
\\ 
&\quad
+
\frac{1}{T}\,
\sum_{k=1}^K \sum_{j=1}^K c(k)  c(j)  (\delta_k - 1) (\delta_j - 1) W_{kj},
\label{eq:trace_scaled_general}
\end{align}
where $W_{kj}$ and $S_{gg, kj}(\theta) $ denote the $[k,j]-$entry of the respective matrix. 
At the $\beta_{0,D}$ and with the optimal weighting $W^* = S_{gg}(\theta_0)^{-1},$ we have
\begin{align}
\label{eq:noise_true_scaled}
\trace\left(W^*S_{gg}(\theta_{0,D})\right)
=
\trace\left(W^*\Delta(D)S_{gg}(\theta_0)\Delta(D)\right)
+
\bs c'\left(\Delta(D)-I_K\right)W^*\left(\Delta(D)-I_K\right)\bs c,
\end{align}
and the partial derivative of the noise term in any scaling direction $d_l$
with $l\in\{1,\ldots,n\}$ is strictly negative, i.e., $\label{eq:directional_derivative_noise}
\left.
\frac{\partial}{\partial d_l}
\frac{1}{T}\trace\left(W^*S_{gg}(\theta_{0,D})\right)
\right|_{D=I_n}
=
-\frac{2}{T}\sum_{k=1}^K k_l
<0.$
\end{proposition}

 A simple way to reduce the scaling bias is to assign larger weights to the variance conditions. A pre-whitened ICA-type procedure represents the limiting case that places infinite weight on the covariance restrictions, thereby removing the scale direction from the optimization problem. However, as discussed below Corollary~\ref{corol:G_a} and shown formally in Appendix~\ref{app:prewhitening}, that comes at the cost of losing optimality of $S_{gg}^{-1}$. 
 An alternative way to eliminate the bias without sacrificing asymptotic efficiency is the continuously updated GMM estimator (CUE) of \cite{han2006gmm}. The SVAR-CUE is given by
\begin{align}
\label{eq:cue_beta}
\hat\beta_T^{\mathrm{cue}}
:=
\argmin_{\beta}
\bar g_T(\hat\alpha_T,\beta)'
\hat W_T(\beta)
\bar g_T(\hat\alpha_T,\beta),
\end{align}
where $\hat W_T(\beta):=\hat S_{gg,T}(\hat\alpha_T,\beta)^{-1}$ and 
$\hat S_{gg,T}(\hat\alpha_T,\beta)$ is an estimator of the second moment matrix of the moment conditions evaluated at the candidate parameter value. With the population analogue $W(\theta)=S_{gg}(\theta)^{-1}$, we have
$
\frac{1}{T}\trace\!\left(W(\theta)S_{gg}(\theta)\right)
=
\frac{1}{T}\trace(I_K)
=
K/T,
$
so the noise term in \eqref{eq: avg finite sample loss 2} is constant in $\theta$ and therefore leads to no bias. The practical difficulty with the CUE is that it requires estimating (and inverting) the covariance matrix $S_{gg}(\theta)$ at every candidate value of $\beta$, which is numerically demanding and can be unstable if the GMM estimator uses many higher-order moment conditions as discussed in the previous section.

The specific form of the scaling distortion in Proposition~\ref{prop:bias} suggests a more targeted correction that avoids continuously re-estimating the full covariance matrix $S_{gg}(\theta)$. For any $\theta=(\alpha_0',\beta')'$, define the population standardization factors 
$d_i(\theta)
:=
\left(\E[e(\theta)_{i,t}^2]\right)^{-1/2}$, 
for $i=1,\ldots,n$ 
and the diagonal matrix
\begin{align}
\label{eq:Lambda_theta}
\Lambda(\theta)
:=
\diag \left(\lambda_1(\theta),\ldots,\lambda_K(\theta)\right),
\qquad
\lambda_k(\theta):=\prod_{i=1}^n d_i(\theta)^{k_i}.
\end{align}
For a fixed positive definite matrix $\widetilde W$, define the scale-updated weighting matrix $W^{\mathrm{su}}(\theta,\widetilde W)
:=
\Lambda(\theta)\widetilde W\Lambda(\theta)$.
By construction, 
at $\theta_{0,D}=(\alpha_0',\beta_{0,D}')'$ with $B(\beta_{0,D})=B_0D$, we have
$d_i(\theta_{0,D})=d_i$, $\lambda_k(\theta_{0,D}) = \delta_k(D)^{-1}$  
and $\Lambda(\theta_{0,D})=\Delta(D)^{-1}.$ 
Hence, a reduction in the variance of the $i$-th innovation, which is equivalent to an increase in the scaling coefficient $d_i$, increases the weights on all moment conditions that involve the $i$-th innovation.
The following proposition shows that the scale-updated weighting matrix exactly offsets the leading scaling distortion.
\begin{proposition}
\label{prop:csue_noise}
Let $\widetilde W$ be any fixed symmetric positive definite $K\times K$ matrix and define
$W^{\mathrm{su}}(\theta,\widetilde W):=\Lambda(\theta)\widetilde W\Lambda(\theta)$, then at the scaled true parameter
$\theta_{0,D}=(\alpha_0',\beta_{0,D}')'$,
\begin{align}
\label{eq:csue_noise_general}
\frac{1}{T}\trace\!\left(
W^{\mathrm{su}}(\theta_{0,D},\widetilde W)\,
S_{gg}(\theta_{0,D})
\right)
&=
\frac{1}{T}\trace\!\left(\widetilde W S_{gg}(\theta_0)\right)
\\ \nonumber
&\quad
+
\frac{1}{T}\,
\bs c'
\bigl(I_K-\Lambda(\theta_{0,D})\bigr)
\widetilde W
\bigl(I_K-\Lambda(\theta_{0,D})\bigr)
\bs c.
\end{align}
In particular, for the asymptotically efficient weight $\widetilde W=W^*=S_{gg}^{-1}$,
\begin{align}
\label{eq:csue_noise_opt}
\frac{1}{T}\trace\!\left(
W^{\mathrm{su}}(\theta_{0,D},W^*)\,
S_{gg}(\theta_{0,D})
\right)
=
\frac{K}{T}
+
\frac{1}{T}\,
\bs c'
\bigl(I_K-\Lambda(\theta_{0,D})\bigr)
W^*
\bigl(I_K-\Lambda(\theta_{0,D})\bigr)
\bs c,
\end{align}
which is minimized at $D=I_n$.
\end{proposition}

Proposition~\ref{prop:csue_noise} shows that the scale-updated weighting matrix removes the leading scaling distortion without continuously updating the covariance matrix $S_{gg}(\theta)$. Unlike the CUE, which updates all entries of $S_{gg}(\theta)$, the scale-updated weighting only adjusts the part of the weighting matrix that is needed to remove the normalization-induced scaling bias.

We can replace $d_i$'s by their sample analogues
$\hat d_{i,T}(\hat\alpha_T,\beta)
:=
\left(
\frac{1}{T}\sum_{t=1}^T e(\hat\alpha_T,\beta)_{i,t}^2
\right)^{-1/2}$ for $i=1,\ldots,n,$ 
and obtain
\begin{align*}
\hat\Lambda_T(\hat\alpha_T,\beta)
:=
\diag\!\left(\hat\lambda_{1,T}(\hat\alpha_T,\beta),\ldots,\hat\lambda_{K,T}(\hat\alpha_T,\beta)\right),
\quad
\hat\lambda_{k,T}(\hat\alpha_T,\beta)
:=
\prod_{i=1}^n \hat d_{i,T}(\hat\alpha_T,\beta)^{k_i}.
\end{align*}
For a fixed weighting matrix $\widetilde W_T$, the continuous scale updating estimator is defined by
\begin{align}
\label{eq:csue_beta}
\hat\beta_T^{\mathrm{csue}}
:=
\argmin_{\beta}
\bar g_T(\hat\alpha_T,\beta)'
\left(
\hat\Lambda_T(\hat\alpha_T,\beta)\,
\widetilde W_T\,
\hat\Lambda_T(\hat\alpha_T,\beta)
\right)
\bar g_T(\hat\alpha_T,\beta).
\end{align}
A natural two-step implementation uses $\widetilde W_T=I_K$ in the first step and
$\widetilde W_T=\hat W_T^*$ in the second step, where $\hat W_T^*$ is a consistent estimator of the asymptotically efficient weight $W^*=S_{gg}^{-1}$ constructed from the first-step estimator as in Section~\ref{sec: estimating s0 and g0}. The feasible construction of $\hat W_T^*$ should, however, be matched to the maintained identifying assumption, i.e., mutual independence~\ref{ass:mutual_indep} or the weaker mean independence~\ref{ass:mean_indep}.

The CSUE remains asymptotically efficient. To see this, suppose that
$\widetilde W_T \cip W$ for some positive definite matrix $W$ and that
$\hat\beta_T \cip \beta_0$. By the unit-variance normalization and consistency,
$\hat d_{i,T}(\hat\alpha_T,\hat\beta_T)\cip 1,$ $\hat\Lambda_T(\hat\alpha_T,\hat\beta_T)\cip I_K,$
and therefore
$\hat\Lambda_T(\hat\alpha_T,\hat\beta_T)\,
\widetilde W_T\,
\hat\Lambda_T(\hat\alpha_T,\hat\beta_T)
\cip W,$
for any $\widetilde W_T \cip W$ by Slutsky's theorem. Hence, the CSUE has the same limiting distribution as the corresponding GMM estimator using the fixed weight $W$. In particular, if $\widetilde W_T=\hat W_T^*$ and $\hat W_T^*\cip W^*=S_{gg}^{-1}$, the two-step CSUE is asymptotically efficient and attains the limit variance in Corollary~\ref{corol:G_a}.

\section{Monte Carlo Simulation}
\label{sec: Finite sample performance}

This section investigates the finite-sample gains from the proposed asymptotically efficient estimator. We first examine how the structured covariance estimation affects the objective function. We then compare the proposed estimators with popular parametric and semi-parametric approaches to identification by higher-order moments, both under mutually independent shocks and under common volatility.

\subsection{Finite-sample effects of structured weighting}
\label{sec:mc-weighting}

To gauge the effect of estimating the efficient weighting matrix, isolated from the reduced-form estimation uncertainty,  the first experiment evaluates the GMM objective function at $\beta_0$  for alternative weighting schemes. We generate the data $u_t=B_0^{(n)}\varepsilon_t$, for $n\in\{2,4\}$, with impact multiplier given in \eqref{eq:mc-gmm-focused-b0} in Appendix~\ref{sec:mc-dgp-details}. The components of $\varepsilon_t$ follow a  Pearson type distribution with mean zero, variance one, skewness $-1$, and excess kurtosis $4$.  We consider $T\in\{100,200,\ldots,1000\}$ and use 1,000 replications. Appendix~\ref{sec:mc-dgp-details} gives the distributional parameters and discusses the moment conditions of this heavy-tailed design.

\begin{figure}[t]
  \centering
  \includegraphics[width=0.7\linewidth]{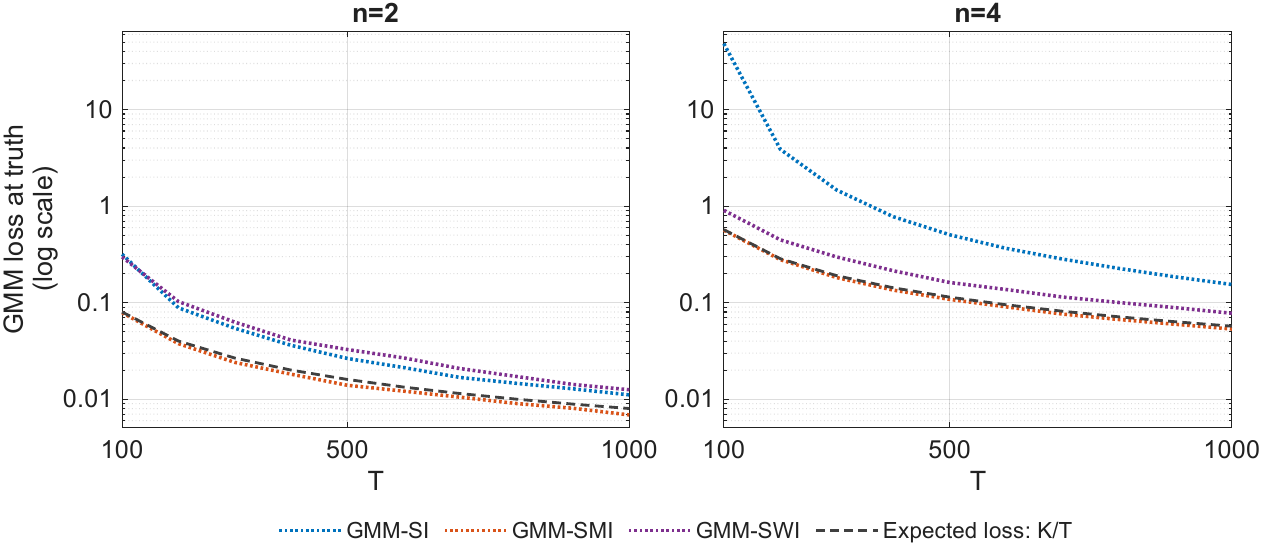}
  \caption{Log GMM loss under alternative weighting schemes.}
  \label{fig:mc-gmm-weighting-loss}
  \par\medskip
  \begin{minipage}{0.95\linewidth}
    \footnotesize
    \textit{Notes:} 
    Dotted curves report Monte Carlo medians of
    $Q_T(\widehat W)$ across 1,000 replications for
    $T\in\{100,200,\ldots,1000\}$. 
    The dashed curve gives the expected loss by using the population optimal weight $K/T$ with $K$ being the number of moment conditions. 
    Figure~\ref{fig:mc-gmm-weighting-loss-full} in Appendix~\ref{sec: appendix mc}
    also reports means, percentile ranges, and the realized oracle losses.
  \end{minipage}
\end{figure}

Figure \ref{fig:mc-gmm-weighting-loss} displays the loss $Q_T(\widehat W)=g_T(\beta_0)'\widehat Wg_T(\beta_0)$ 
where $g_T(\beta_0)$ contains the sample second- to fourth-order moment conditions at the true impact matrix. GMM-SI uses the conventional estimator for the asymptotically efficient weighting in \eqref{eq: S SI}. GMM-SMI exploits mutual independence to replace joint co-moments by products of marginal moments as in \eqref{eq: S SIMI}. GMM-SWI exploits the covariance restrictions implied by mean independence in \eqref{eq: S SWI}. We also display as a benchmark the loss from an infeasible oracle estimator using the population weight $S_{gg}^{-1}$.

The simulation shows that the conventional estimator for the asymptotically
efficient weighting matrix is inadequate for approximating the true asymptotically efficient weighting matrix, especially for large SVAR models. The magnitude of this improvement should be read on the logarithmic vertical axes.
Specifically,  at $n=4$ and $T=100$, the median loss falls from approximately $50$ for GMM-SI to $0.56$ for GMM-SMI, a reduction of almost two orders of magnitude. GMM-SWI also lowers the median loss markedly, to approximately $0.90$. The difference remains visible at $T=1000$, where the GMM-SI median is still about three times the GMM-SMI median and twice the GMM-SWI median. Thus, the difficulty of estimating the weighting matrix is not confined to the smallest sample size. 
The comparison between GMM-SI and GMM-SMI is particularly informative. Although both use the same identifying moments, the structured weight produces losses much closer to the infeasible benchmark.
%If an analyst intends to compare the GMM estimator with identification approaches that rely on mutual independence, such a comparison shall be conduced on equal footing, so that GMM-SMI, rather than GMM-SI, should be used in the comparison. As we illustrate next, improvement in estimation of the efficient weight translates directly to a more precise estimator for $\beta$ and sharper inference for the structural IRFs.

 \subsection{Comparison with alternative higher-order estimators}
\label{sec:mc-comparison}

We next examine whether the improvement in feasible GMM makes its performance comparable to benchmark approaches that require mutual independence, and how the comparison changes when this assumption is violated. We consider a four-dimensional VAR(1) with the same structural impact multiplier as in the previous experiment
\begin{equation}
 y_t=\begin{bmatrix}
 0.42&0&0.08&0\\
 0.08&0.18&0.05&0\\
 0.06&0.05&0.38&0.04\\
 0.04&0.02&0.07&0.32
 \end{bmatrix}y_{t-1}+\begin{bmatrix}
     2.8 &  0   & 0   & 0 \\
    -0.6 &  2.2 & 0   & 0 \\
    -1.6 &  1.1 & 2.6 & 0 \\
     1.2 & -0.9 & 0.7 & 2.4
  \end{bmatrix}\varepsilon_t,
 \label{eq:mc-lag-matrix}
\end{equation}
and consider $T\in\{250,500,1000\}$. 
We first consider the scenario of independent shocks, in which the shocks follow the same Pearson distribution as in Section~\ref{sec:mc-weighting}. 
In the second design, some shocks exhibit a common stochastic volatility, i.e., $\varepsilon_{it}=\sqrt{W_{it}}z_{it}$, where the $z_{it}$ are independent draws from a two-component normal mixture and are independent of the volatility paths.
Each $W_{it}$ follows a two-state Markov chain, in which the staying probabilities in the lower and higher volatility states, characterized by standard-deviation multipliers of approximately $0.109$ and $1.818$, are 0.78 and 0.48, respectively. The stationary probability of the high volatility state is $0.30$. The Markov switching volatility design is particularly relevant in many financial and macroeconomic models, in which volatilities of certain shocks co-move, for instance, during transitions between peaceful growth periods (e.g., the \textit{Great Moderation}) and periods of economic downturns and/or financial distress.
Shocks 1 and 2 share their entire volatility path, while shocks 3 and 4 have independent paths. All shocks have mean zero, variance one, skewness approximately $-1.8$, and excess kurtosis $7$. The mixture parameters and transition probability matrix are given in Appendix~\ref{sec:mc-dgp-details}, where we also report the shocks' higher-order moments and highlight that the symmetric cokurtosis condition $\E[\varepsilon_{1t}^2\varepsilon_{2t}^2]=1$ is violated.

 \begin{figure}[t]
  \centering
  \includegraphics[width=\linewidth]{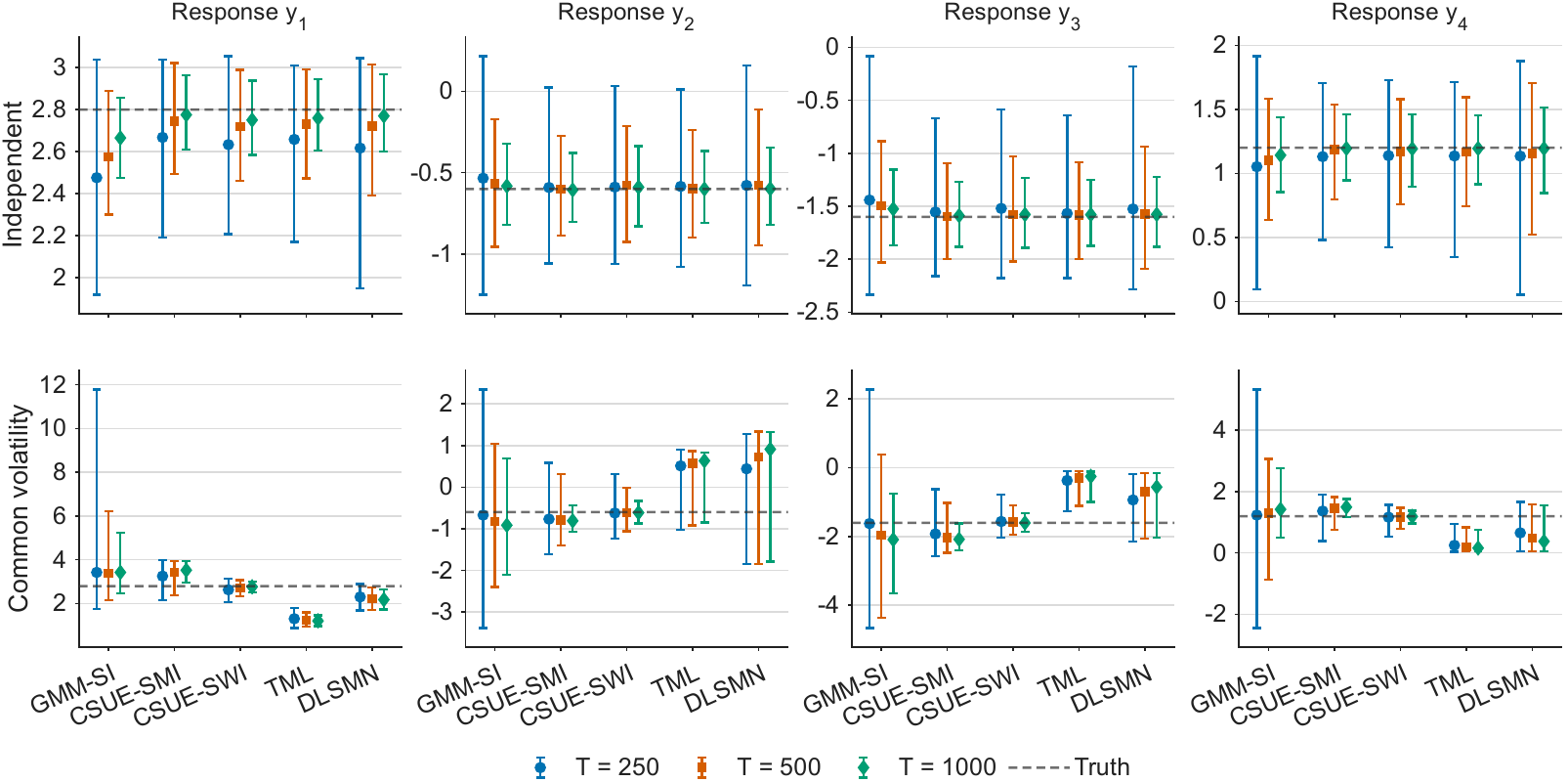}
  \caption{Distributions of estimated impact responses under independent shocks and common volatility.}
  \label{fig:mc-cv-dlsmn-saddle-shock1-h0-ranges-90}
  \par\medskip
  \begin{minipage}{0.98\linewidth}
    \footnotesize
    \textit{Notes:} Markers report Monte Carlo medians and vertical bars report the 5th and 95th percentiles across 1000 replications. The top row is the independent-shock benchmark and the bottom row is the common-volatility design, columns give the responses of $y_1,\ldots,y_4$ to structural shock 1 at horizon $h=0$.  Colors and marker shapes distinguish $T\in\{250,500,1000\}$. Dashed horizontal lines show the true responses. 
  \end{minipage}
\end{figure}
In every replication, we estimate the reduced-form VAR by OLS and compare five estimators of the impact responses. GMM-SI uses all second- to fourth-order moment conditions and the conventional two-step estimator for the efficient weighting matrix  based on the sample covariance matrix in \eqref{eq: S SI}. 
CSUE-SMI uses the same moment conditions and the continuous scale update from  Section \ref{sec: bias}, and the   second-step weighting matrix is estimated under the assumption of mutual independence as in Equation \eqref{eq: S SIMI}.
CSUE-SWI drops the six symmetric cokurtosis restrictions and combines the mean-independence covariance estimator in \eqref{eq: S SWI} with the continuous scale update from  Section \ref{sec: bias}.
We compare the moment based estimators with two likelihood-based estimators:  the TML, which uses a Student-$t$ pseudo-likelihood following \citet{lanne2017identification}, and the discrete location scale mixture of normals (DLSMN) estimator proposed by \cite{fiorentini2023discrete}. Compared to TML, which uses a fixed density function, the recently proposed DLSMN estimator allows the density in the likelihood function to be flexibly adjusted given the data. In specific, the density is specified by a finite normal mixture and  the structural parameter $\beta$ is jointly estimated with the shape and weights of the mixture distribution. The DLSMN approach is consistent regardless of the true distribution of the shocks and has demonstrated remarkable simulation performance in recent studies \citep{fiorentini2023discrete, hafner2025kml}. However, it does require the strong mutual independence Assumption~\ref{ass:mutual_indep}.

Figure~\ref{fig:mc-cv-dlsmn-saddle-shock1-h0-ranges-90}  reports the medians and central 90\% ranges of the estimated impact responses to the first shock. Under independent shocks, both CSUE  estimators substantially reduce dispersion relative to GMM-SI. At $T=250$, their central ranges are about one quarter to one third narrower across the four responses. This makes the performance of the proposed CSUE estimators comparable to TML and DLSMN, often with narrower central ranges. These findings contrast with the weaker relative performance of GMM reported, for example, by \cite{hafner2025kml}. In the present design, exploiting the maintained independence restrictions in the weight and addressing the scale distortion makes GMM competitive with the likelihood-based alternatives.

Under common volatility, the differences among these estimators show up in the bias and the dispersion. As the mutual independence assumption is violated in this case, both TML and DLSMN exhibit pronounced bias. For example, their median responses of $y_2$ have the wrong sign even at $T=1000$. Their sampling ranges become narrower around values away from the truth. CSUE-SMI is also affected by imposing mutual independence, particularly for the responses of $y_1$ and $y_3$. In contrast, the CSUE-SWI medians remain close to the true impact vector and its sampling ranges contract around it.

Figure~\ref{fig:mc-bootstrap-cv-dlsmn-saddle-shock1-h0} in Appendix~\ref{sec: appendix mc} gives the corresponding coverage and average length of pointwise 90\% intervals obtained from a residual-based moving block bootstrap \citep{BJT_MBB2016}.\footnote{Appendix~\ref{sec:mc-implementation} records the  bootstrap construction.} Under independent shocks, without systematic undercoverage, the CSUE intervals are substantially shorter than those of GMM-SI, with average lengths reduced by roughly one half at $T=250$, and are broadly comparable to the likelihood-based alternatives.
Under common volatility, however, the bias of TML and DLSMN is accompanied by severe undercoverage, which becomes particularly pronounced as $T$, with coverage often falling into single digits. CSUE-SWI provides a much more favorable balance of coverage and interval length. The high coverage of GMM-SI is accompanied by extremely long intervals.

\section{Oil-market application}
\label{sec:Application}

This section illustrates the effect of structured weighting  and continuous scale updating in an oil-market SVAR.
We compare the conventional GMM-SI estimator with a second-step covariance estimator based on  Equation  \eqref{eq: S SI} 
with the proposed CSUE-SWI estimator, which uses only mean independence to estimate the second-step weighting matrix based on Equation \eqref{eq: S SWI} with a continuous scale update from Section \ref{sec: bias}. 

%
%\subsection{Data and specification}

The application belongs to the oil-market literature initiated by \citet{kilian2009not}, which distinguishes oil-supply, aggregate-activity, and oil-specific-demand shocks, and its extension to stock prices by \citet{kilian2009impact}. A conventional recursive identification that places stock prices last rules out an immediate response of oil-market variables to stock-market innovations. This restriction is consequential when stock prices contain forward-looking information about future economic activity or oil demand, see \citet{keweloh2026uncertain}. %\citet{keweloh2026uncertain} instead use non-Gaussian higher moments to identify the shocks while treating the short-run zero restrictions as uncertain. Their estimates attribute a non-negligible share of oil-price variation to a stock-market-information shock.

We retain the four-variable system of \citet{keweloh2026uncertain} and extend the monthly sample to January 1974--September 2025. The variables are world crude-oil production, $q_t$, global industrial production, $y_t^{\mathrm{IP}}$, the real oil price, $p_t$, and a real U.S. stock-price index, $s_t$, with $q_t$ measured as 100 times the logarithm of production. We estimate an unrestricted VAR(12). The  structural shocks are labeled oil supply ($\varepsilon_t^S$), aggregate demand ($\varepsilon_t^Y$), oil-specific demand ($\varepsilon_t^D$), and stock-market information ($\varepsilon_t^{SM}$), based on the response patterns and forecast variance contributions discussed below.

Both the GMM-SI and the CSUE-SWI estimators use the second- to fourth-order moment conditions derived under the mutual and weaker mean independence assumptions, respectively.
In particular, CSUE-SWI omits the six symmetric cokurtosis restrictions $\E[\varepsilon_{it}^2\varepsilon_{jt}^2]=1$, $i\neq j$. Therefore, potential dependence in the second-order moments of the shocks (i.e., common stochastic volatility) is permitted.
The estimator GMM-SI uses the sample covariance of the retained moment functions as a fixed second-step weight.
The estimator CSUE-SWI constructs this covariance under mutual mean independence and applies the continuous scale update developed in Section~\ref{sec: bias}. 

\begin{figure}[t]
    \centering
    \includegraphics[width=0.5\textwidth]{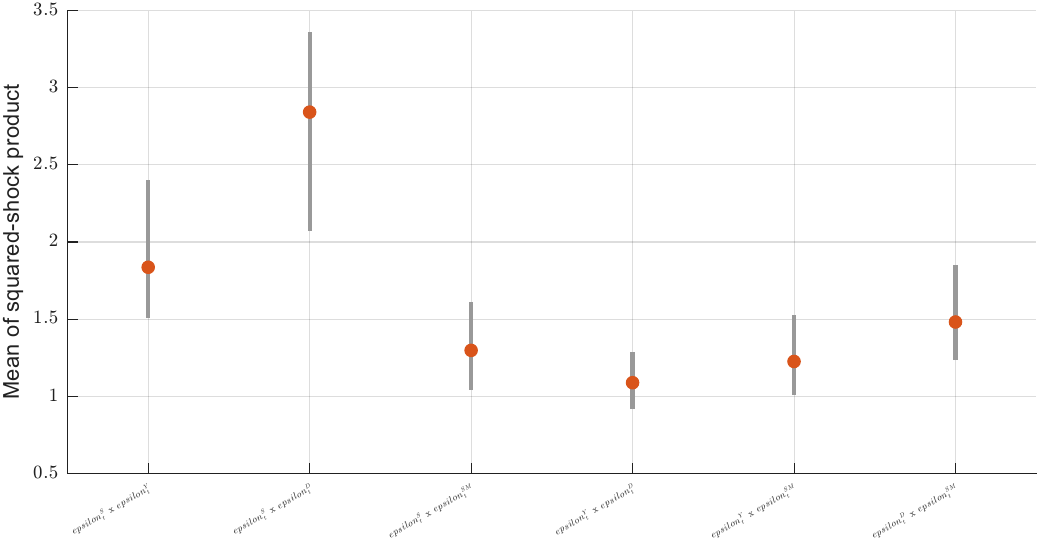}
    \caption{Dependence in squared CSUE-SWI structural shocks. }
    \label{fig:oil-dependence}
    % Bars report $\widehat D_{ij}=\frac{1}{T}\sum_{t=1}^{T}
    % \widehat\varepsilon_{it}^{2}\widehat\varepsilon_{jt}^{2}$ with are pointwise median-centered 68\% moving-block-bootstrap intervals.  
    \begin{minipage}{0.98\linewidth}
    \footnotesize
    \textit{Notes:} 
    Markers report $\widehat D_{ij}$ and vertical bars are pointwise median-centered 68\% moving-block-bootstrap intervals. The horizontal line at one is the value implied by independence and unit variances. Shock labels are oil supply ($S$), aggregate demand ($Y$), oil-specific demand ($D$), and stock-market information ($SM$).
  \end{minipage}
\end{figure}

The estimated co-moments of the shocks provide the motivation for allowing dependence in their variances. Figure~\ref{fig:oil-dependence} reports the squared-shock dependence statistic $\widehat D_{ij}=\frac{1}{T}\sum_{t=1}^{T}
    \widehat\varepsilon_{it}^{2}\widehat\varepsilon_{jt}^{2}.$ 
Under unit-variance normalization, $D_{ij}=1$ under independence. However, all six estimates exceed one, and five of the six pointwise 68\% intervals exclude one. The exception is the aggregate--oil-demand pair. Especially large co-movements are observed for variances in oil supply and oil-specific demand, which align with the magnitude dependence emphasized by \citet{keweloh2026uncertain}. As opposed to the aggregate demand ($\varepsilon_t^Y$), which reflects the demand for ``flow'' of oil, the oil-specific demand is intended to capture the precautionary demand for ``stock'' of oil \citep[see related discussions in][and output responses to $\varepsilon_t^D$ below]{Kilian_Murphy_2014, Anzuini_2015, Wang2026QE}. Hence, the detected dependence pattern is consistent with the notion that the precautionary demand for oil may exhibit increased volatility amidst heightened uncertainty surrounding crude oil supply due to factors such as rising geopolitical tensions, warfare, and sanctions, as recently put forward by \cite{HW_2024_HCKML}.
Appendix~\ref{app:application} reports the shocks' skewness and excess kurtosis and examines the consequences of imposing mutual independence instead.

%Inference is based on 3,000 valid moving-block-bootstrap replications for each estimator. Each replication resamples the complete vector of reduced-form residuals in blocks of length eight, simulates the VAR from the original 12 initial observations, and re-estimates both the reduced form and the structural model. The reported intervals are pointwise 68\% intervals centered on the point estimate using distances between the bootstrap median and the 16th and 84th percentiles. They are not simultaneous bands, and the solid curves in the impulse-response figures are point estimates rather than bootstrap medians.

%\subsection{Dynamic effects and empirical precision}

\begin{figure}[h] 
    \centering
    \includegraphics[width=0.90\textwidth]{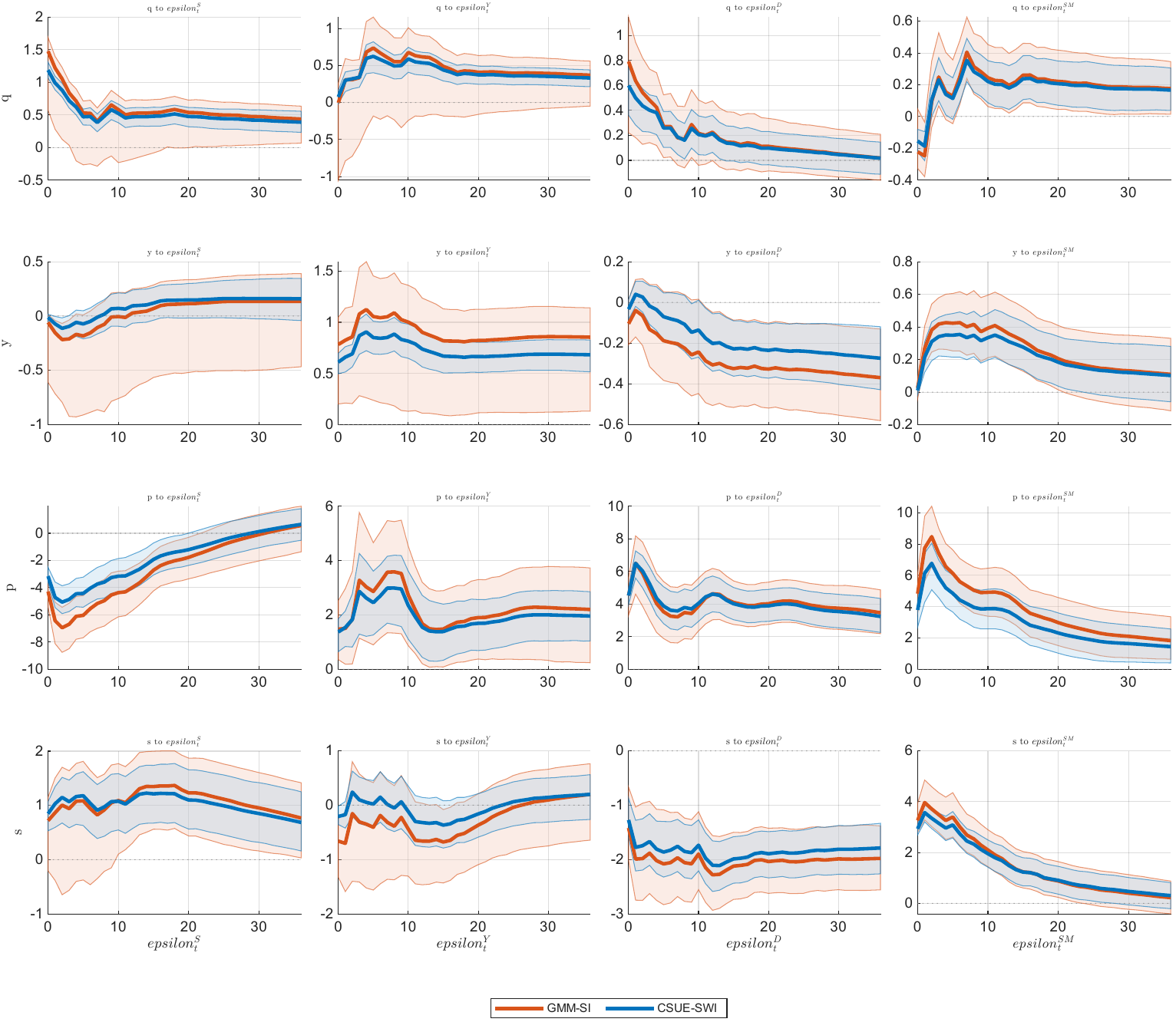}
    \caption{Impulse responses from GMM-SI and CSUE-SWI. }
    \label{fig:oil-irfs}
    \begin{minipage}{0.98\linewidth}
    \footnotesize
    \textit{Notes:} 
    Rows index responses, columns index structural shocks, with horizons from zero to 36 months. Solid curves are point estimates and shaded regions are pointwise median-centered 68\% intervals from 3,000 moving-block-bootstrap replications with block length eight.
  \end{minipage}
\end{figure}

Figure~\ref{fig:oil-irfs} compares the GMM-SI and CSUE-SWI responses. A positive supply shock raises oil production and lowers the real oil price on impact. The aggregate demand shock has the largest contemporaneous effect on global industrial production and is followed by increases in oil production and the real oil price.
The oil-specific-demand shock raises production and the oil price, with a small and imprecisely estimated impact effect on activity and a subsequent decline in activity over the medium run.
A positive stock-market-information shock immediately raises stock prices and the real oil price. Oil production initially falls and subsequently rises, while the activity response has a positive lower 68\% confidence bound from month one through month 23. These response patterns are comparable with those provided in previous work \citep[see, e.g.,][]{Kilian_Murphy_2014, keweloh2026uncertain}.

The main difference shows up in how precisely these structural IRFs are estimated.
Specifically, the proposed CSUE-SWI estimator yields narrower pointwise intervals throughout the displayed responses and horizons than those obtained with conventional GMM implementation.
The estimation efficiency and thus sharper inference affect the empirical conclusions as well. For example, the stock-price response to a positive supply shock has a positive impact estimate under both procedures in terms of point estimate. However, the confidence interval provided by the CSUE-SWI  clearly excludes zero, whereas the wider GMM-SI interval includes it and indicates an  insignificant stock market response to oil supply shocks.

Table~\ref{tab:oil-fevd} reports the CSUE-SWI forecast-error variance decomposition at a 12 months  horizon. Oil production is explained mainly by supply shocks, followed by aggregate demand  and oil-specific demand. Global activity is dominated by aggregate demand shocks, with stock-market information accounting for 13\%. The forecast uncertainty of oil price is spread across stock-market information, oil-specific demand, and supply. Stock-market information accounts for 65\% of stock-price forecast uncertainty, while oil-specific demand and supply account for 26\% and 10\%, respectively.  Oil production and real activity decompositions are close to the corresponding ridge estimates in \citet{keweloh2026uncertain}. The main difference concerns the real oil price. Their analysis assigns 18\% of forecast-error variance of oil price to stock-market information and 45\% to oil-specific demand, compared with 35\% and 30\% reported here. It strengthens the conclusion that forward-looking stock prices contain relevant information about future oil-market conditions relevant for oil-price movements, which a recursive ordering that excludes their immediate oil-market effects obscures. 

\begin{table}[t]
    \centering
    \caption{CSUE-SWI forecast-error variance at a 12 months horizon. Cells report  point estimates with pointwise median-centered 68\% moving-block-bootstrap intervals.}
    \label{tab:oil-fevd}
    \begin{tabular}{lcccc}
        \toprule
        Response & $\varepsilon_t^{S}$ & $\varepsilon_t^{Y}$ & $\varepsilon_t^{D}$ & $\varepsilon_t^{SM}$ \\
        \midrule
      q & $\underset{0.44/0.66}{0.55}$ & $\underset{0.18/0.36}{0.26}$ & $\underset{0.06/0.22}{0.13}$ & $\underset{0.02/0.11}{0.05}$ \\
y & $\underset{-0.01/0.04}{0.01}$ & $\underset{0.74/0.93}{0.85}$ & $\underset{0.00/0.05}{0.01}$ & $\underset{0.06/0.22}{0.13}$ \\
p & $\underset{0.15/0.36}{0.25}$ & $\underset{0.04/0.16}{0.09}$ & $\underset{0.20/0.45}{0.30}$ & $\underset{0.23/0.47}{0.35}$ \\
s & $\underset{0.05/0.17}{0.10}$ & $\underset{-0.01/0.03}{0.00}$ & $\underset{0.14/0.39}{0.26}$ & $\underset{0.51/0.78}{0.65}$ \\
        \bottomrule
    \end{tabular}
\end{table}

\section{Conclusion}
\label{sec:conclusion}

This paper establishes conditions for the asymptotic optimality of the conventional SVAR-GMM weighting and develops a feasible implementation through which efficiency gains become relevant in finite samples. Estimating only autoregressive slopes in the first step yields adaptivity under mild conditions. To make the optimal weighting feasible and stable, we use the same independence assumptions in estimation of the moment covariance as in identification. Under mutual independence, a large collection of high-order co-moments is replaced by a small set of lower-order marginal moments. The weaker mean independence assumption permits less simplification but accommodates shocks that share volatility. Continuous scale updating removes the leading scaling incentive by adjusting only the relevant part of the weighting, while retaining asymptotic efficiency. The simulations demonstrate that the poor finite-sample performance of conventional GMM can be explained by its implementation. Structured covariance estimation and continuous scale updating bring its performance to the same level as competing likelihood-based estimators under independent shocks. When shocks share common volatility, the mean-independence estimator remains centered close to the true impact responses, whereas the competing estimators exhibit pronounced bias. In the oil-market application, the resulting precision gains sharpen the evidence regarding stock market responses to unexpected oil supply disruptions.

 \bibliographystyle{chicago}
 \bibliography{literatur}

	\newpage 
	
 \appendix

\section{Proof}
 \label{app: proof}

\begin{proof}[Proof of Proposition~\ref{thm:asymptotics}]
For the first step, the normal equations imply
\[
0=\frac1T\sum_{t=1}^T m_t(\hat\alpha_T)
=
\frac1T\sum_{t=1}^T m_t(\alpha_0)
-
\left(\frac{ZZ'}{T}\otimes I_n\right)(\hat\alpha_T-\alpha_0),
\]
because $\partial m_t(\alpha)/\partial\alpha'=-(Z_{t-1}Z_{t-1}')\otimes I_n$ is constant in $\alpha$. Therefore
\[
\sqrt{T}(\hat\alpha_T-\alpha_0)
=
\hat M_T^{-1}\frac1{\sqrt T}\sum_{t=1}^T m_t(\alpha_0),
\]
where $\hat M_T =\left(\frac{ZZ'}{T} \otimes I_n\right)$. Expanding the first-order condition to the second-step estimator yields
	\begin{align*}
	\sqrt{T}(\hat \beta_T - \beta_0) &=
	- \left[\left(\frac{1}{T}\sum_{t=1}^T \frac{\partial g(y_t, \hat \alpha_T, \beta^*) }{\partial \beta'} \right)'
	\hat W_T
	\left(\frac{1}{T}\sum_{t=1}^T \frac{\partial g(y_t, \hat \alpha_T, \beta^*) }{\partial \beta'} \right)
	\right]^{-1} \times \\
	& \hspace{.7cm} \left(\frac{1}{T}\sum_{t=1}^T \frac{\partial g(y_t, \hat \alpha_T, \beta^*) }{\partial \beta'} \right)'
	\hat W_T \frac{1}{\sqrt{T}}\sum_{t=1}^T g(y_t, \hat \alpha_T, \beta_0) 
	\end{align*}
	where $\beta^*$ is a mean value between $\hat \beta_T $ and $\beta_0$ and
	\begin{align*}
		\frac{1}{\sqrt{T}}\sum_{t=1}^T g(y_t, \hat \alpha_T, \beta_0) &= \frac{1}{\sqrt{T}}\sum_{t=1}^T g(y_t,  \alpha_0, \beta_0) + \left(\frac{1}{T}\sum_{t=1}^T \frac{\partial g(y_t, \alpha^*, \beta_0) }{\partial \alpha'} \right) \sqrt{T}(\hat \alpha_T - \alpha_0).
	\end{align*} 
    where $\alpha^*$ is a mean value between $\hat\alpha_T$ and $\alpha_0$.

    Finally, by Lemma~\ref{lem:LLN_ULLN}(i), $\hat M_T \cip M,$ 
while Lemma~\ref{lem:LLN_ULLN}(ii), Assumption~\ref{ass:beta}(ii), and consistency of $(\hat\alpha_T,\hat\beta_T)$ imply that
\[
\frac1T\sum_{t=1}^T \frac{\partial g(y_t,\alpha^*,\beta_0)}{\partial\alpha'}  \cip \E[G_{\alpha, t}(\theta_0)] = G_\alpha,
\quad
\frac1T\sum_{t=1}^T \frac{\partial g(y_t,\hat\alpha_T,\beta^*)}{\partial\beta'} \cip \E[G_{\beta, t}(\theta_0)] = G_\beta.
\]
Since $W$ is positive definite and $\rank G_\beta = n^2$  by \ref{ass:beta}, $G_\beta'WG_\beta$ is invertible, and hence 
\begin{align*}
    \hat J_T^{-1}
&:=
\begin{pmatrix}
\hat M_T^{-1} & 0\\
-\hat H_T \left(\frac1T\sum_{t=1}^T \frac{\partial g(y_t,\alpha^*,\beta_0)}{\partial\alpha'}\right)\hat M_T^{-1} & -\hat H_T
\end{pmatrix}\\
&\cip
J^{-1}
:= 
\begin{pmatrix}
M^{-1} & 0\\
-(G_\beta'WG_\beta)^{-1}G_\beta'WG_\alpha M^{-1}
&
-(G_\beta'WG_\beta)^{-1}G_\beta'W
\end{pmatrix}.
\end{align*}
where 
\begin{align*}
    \hat H_T &:= \left[ \left(\frac1T\sum_{t=1}^T \frac{\partial g(y_t,\hat\alpha_T,\beta^*)}{\partial\beta'}\right)' \hat W_T \left(\frac1T\sum_{t=1}^T \frac{\partial g(y_t,\hat\alpha_T,\beta^*)}{\partial\beta'}\right) \right]^{-1} \times \\
	& \hspace{.7cm} 
   \left(\frac1T\sum_{t=1}^T \frac{\partial g(y_t,\hat\alpha_T,\beta^*)}{\partial\beta'}\right)' \hat W_T \cip H := (G_\beta'WG_\beta)^{-1}G_\beta'W
\end{align*}
Let $U_t = (m_t(\alpha_0)', g_t(\theta_0)')'$, we have 
\[
\sqrt T
\begin{pmatrix}
\hat\alpha_T-\alpha_0\\
\hat\beta_T-\beta_0
\end{pmatrix}
= J^{-1}
\frac1{\sqrt T}\sum_{t=1}^T U_t + o_p(1)
\ \Rightarrow \
\mc N \left(0,\ J^{-1}S_{uu}(J^{-1})'\right).
\]
by Slutsky's theorem and Lemma~\ref{lem:CLT_iid},
where the asymptotic covariance matrix is
\begin{align*}
    V_\theta &= J^{-1}S_{uu}(J^{-1})'\\
    &= \begin{pmatrix}
M^{-1} & \bs 0_{n^2 p \times K}\\
-HG_\alpha M^{-1}
&
-H
\end{pmatrix}
\begin{pmatrix}
S_{mm} & S_{mg}\\
S_{mg}' & S_{gg}
\end{pmatrix}
\begin{pmatrix}
M^{-1} & -M^{-1}G_\alpha' H'\\
\bs 0_{K \times n^2p}
&
-H'
\end{pmatrix}\\
&= \begin{pmatrix}
\Gamma^{-1} \otimes \Sigma_u 
&  - \left[(\Gamma^{-1} \otimes \Sigma_u ) G_\alpha' + M^{-1}S_{mg} \right]  H'\\
-H \left[G_\alpha (\Gamma^{-1} \otimes \Sigma_u ) + S_{mg}' M^{-1} \right] 
& H \left[S_{gg} 
+ G_\alpha M^{-1} S_{mg}
+ S_{mg}' M^{-1} G_\alpha'
+ G_\alpha (\Gamma^{-1} \otimes \Sigma_u ) G_\alpha' \right] H'
\end{pmatrix},
\end{align*}
since $M^{-1} S_{mm} M^{-1} = M^{-1}(\Gamma\otimes \Sigma_u)M^{-1} = \Gamma^{-1} \otimes \Sigma_u$.
To see the influence function for the GMM estimator of the structural parameters, note that  
\begin{align*}
    \sqrt{T}(\hat\beta_T-\beta_0)
&=
-(G_\beta'WG_\beta)^{-1}G_\beta'W
\left(
\frac1{\sqrt T}\sum_{t=1}^T g_t(\alpha_0, \beta_0)
+
G_\alpha \sqrt{T}(\hat\alpha_T-\alpha_0)
\right)
+o_p(1)\\
&= -(G_\beta'WG_\beta)^{-1}G_\beta'W \frac1{\sqrt T}\sum_{t=1}^T \left(g_t(\alpha_0, \beta_0) + G_\alpha M^{-1} m_t(\alpha_0) \right) + o_p(1)\\
&=: -(G_\beta'WG_\beta)^{-1}G_\beta'W \frac1{\sqrt T}\sum_{t=1}^T g_t^{\perp}(\alpha_0, \beta_0)  + o_p(1),
\end{align*}
where 
\[
S_\perp := \E[g_t^{\perp}(\alpha_0, \beta_0) g_t^{\perp}(\alpha_0, \beta_0)'] = S_{gg}  + G_\alpha M^{-1} S_{mg} + S_{mg}' M^{-1} G_\alpha' + G_\alpha (\Gamma^{-1} \otimes \Sigma_u ) G_\alpha',
\]
This completes the proof.
\end{proof}

\begin{proof}[Proof of Proposition~\ref{prop:bias}]
\eqref{eq:trace_scaled_general} follows from Lemma~\ref{lem:scale_noise}, written in compact notation (see the proof of the Lemma for details):
\begin{align*}
\frac{1}{T}\trace\bigl(W S_{gg}(\theta_D)\bigr)
&=
\frac{1}{T}\trace\bigl(W\Delta(D)S_{gg}(\theta)\Delta(D)\bigr)
+
\frac{2}{T} \bs c'\bigl(\Delta(D)-I_K\bigr)W\Delta(D)\E[g_t(\theta)]
\\ 
&\quad
+
\frac{1}{T}\bs c'\bigl(\Delta(D)-I_K\bigr)W\bigl(\Delta(D)-I_K\bigr) \bs c.
\end{align*}
At $\theta=\theta_0$, we have $\E[g_t(\theta_0)]=\bs 0,$ 
so the cross term vanishes and \eqref{eq:noise_true_scaled} follows immediately.

To compute the partial derivative of the noise term, define for each $l=1,\ldots,n$ the diagonal matrix
$
H_l := \diag(k_l:\,k=1,\ldots,K),
$
that is, the $k$-th diagonal element of $H_l$ equals the exponent of the $l$-th innovation in the $k$-th moment condition in \eqref{eq: f}. Then, at $D = I_n = \diag(1,\ldots,1)$, we have
\[
\left.
\frac{\partial \Delta(D)}{\partial d_l}
\right|_{D=I_n}
=
-H_l.
\]
Differentiating \eqref{eq:noise_true_scaled} yields
\begin{align*}
\left.
\frac{\partial}{\partial d_l}
\trace\bigl(W^*S_{gg}(\theta_{0,D})\bigr)
\right|_{D=I_n}
&=
\left.
\frac{\partial}{\partial d_l}
\trace\bigl(W^*\Delta(D)S_{gg}(\theta_0)\Delta(D)\bigr)
\right|_{D=I_n}
\\
&=
\trace\bigl(W^*(-H_l)S_{gg}(\theta_0)\bigr)
+
\trace\bigl(W^*S_{gg}(\theta_0)(-H_l)\bigr),
\\
&= -2\trace(H_l)
=
-2\sum_{k=1}^K k_l.
\end{align*}
where first line follows from the fact that the derivative of the second term of \eqref{eq:noise_true_scaled} is zero at $D=I_n$ as it is quadratic in $\Delta(D)-I_K$,
while the second equality is due to $ W^*S_{gg}(\theta_0)=S_{gg}(\theta_0)W^*=I_K$. This completes the proof.
\end{proof}

\begin{proof}[Proof of Proposition~\ref{prop:csue_noise}]
At $\theta_{0,D}$, since $\E[g_t(\theta_0)]=\bs 0$, by Lemma~\ref{lem:scale_noise}, we have
\[
S_{gg}(\theta_{0,D})
=
\Delta(D)S_{gg}(\theta_0)\Delta(D)
+
\left(\Delta(D)-I_K\right)\bs c\bs c'
\left(\Delta(D)-I_K\right).
\]
Since $\Lambda(\theta_{0,D})=\Delta(D)^{-1}$, we have
\begin{align*}
\trace\!\left(
W^{\mathrm{su}}(\theta_{0,D},\widetilde W)\,
S_{gg}(\theta_{0,D})
\right)
&=
\trace\!\left(
\Lambda(\theta_{0,D}) \widetilde W\Lambda(\theta_{0,D})
\Delta(D) S_{gg}(\theta_0)\Delta(D)
\right)\\
&\quad +
\trace\!\left(
\Lambda(\theta_{0,D})\widetilde W\Lambda(\theta_{0,D})
(\Delta(D)-I_K)\bs c\bs c'(\Delta(D)-I_K)
\right)
\\
&=
\trace\!\left(\widetilde W S_{gg}(\theta_0)\right)
+
\bs c'(I_K-\Lambda(\theta_{0,D}))\widetilde W(I_K-\Lambda(\theta_{0,D}))\bs c,
\end{align*}
which proves the first part. The second part follows by noticing that at $\widetilde W=W^*=S_{gg}^{-1}$, $\trace\!\left(W^*S_{gg}(\theta_0)\right)=K$. Since $W^*$ is positive definite, the second term is nonnegative and equals zero at $D=I_n$. 
\end{proof}

% -------------------------------------------- &
% -------------------------------------------- &

\subsection{Lemmas}
\begin{lemma}\label{lem:CLT_iid}
Let $U_t = (m_t(\alpha_0)', g_t(\theta_0)')'$, under Assumptions of Proposition~\ref{thm:asymptotics}, 
    \[
\frac1{\sqrt T}\sum_{t=1}^T U_t
\ \Rightarrow \
\mc N(0,S_{uu}),
\qquad
S_{uu}=
\begin{pmatrix}
S_{mm} & S_{mg}\\
S_{mg}' & S_{gg}
\end{pmatrix},
\]
where
$
S_{mm}=\Gamma\otimes \Sigma_u, 
\;
S_{gg}=\E[g(\varepsilon_t)g(\varepsilon_t)'],
\;
S_{mg} = (\E[Z_{t-1}]\otimes I_n)\,B_0\,\E[\varepsilon_t g(\varepsilon_t)'].
$
Hence, if $\E[Z_{t-1}]=0$, 
we have
\[
S_{uu}=
\begin{pmatrix}
S_{mm} & \bs 0_{n^2 p \times K}\\
\bs 0_{K \times n^2 p} & S_{gg}
\end{pmatrix}.
\]
\end{lemma}

% -------------------------------------------- &
% -------------------------------------------- &
\begin{proof}[Proof of Lemma~\ref{lem:CLT_iid}]
    Let $\lambda \in \R^{n^2p + K}$ with $\lambda'\lambda = 1$, define $V_{t} := \lambda'U_{t}$ and by Cram\'er-Wold device, it suffices to show 
	\[
		\frac{1}{\sqrt{T}}\sum_{t=1}^T V_t \ \Rightarrow \ \mc N (0, s^2_\lambda),
        \qquad
        s^2_\lambda := \lambda'S_{uu}\lambda
	\]
    Since the map $\varepsilon_t \mapsto u_t = B\varepsilon_t$ is a bijection with measurable inverse by~\ref{ass:base}(i),  $\F_t := \sigma(\varepsilon_s:s\le t) = \sigma(u_s:s\le t)$ by Thm.\ 10.4 in \cite{Davidson2021}. Hence, $u_t$ is $\F_{t}$-measurable with $\E[u_t \mid \F_{t-1}] = B\E[\varepsilon_t \mid \F_{t-1}] = 0$, and
    $m_t$ is $\F_{t}$-measurable with $\E[m_t \mid \F_{t-1}] = (Z_{t-1} \otimes I_n) \E[u_t \mid \F_{t-1}]  = 0$. Moreover, by Assumptions~\ref{ass:base}(ii) and~\ref{ass:beta}(i), $ \E[g_t(\theta_0)\mid \F_{t-1}] = \E[g(\varepsilon_t)\mid \F_{t-1}] = \E[g(\varepsilon_t)] = 0$. Therefore, $\{U_t, \F_t\}$ and hence $\{V_t, \F_t\}$ are MDS.

    Define $X_{T,t} := V_t/(\sqrt{T} s_\lambda)$, note that $\sum^T_{t=1} \E[X_{T,t}^2] = 1$ by construction. Next, we verify conditions in Thm.\ 25.3 in \cite{Davidson2021}:
    (a) $\sum^T_{t=1} X_{T,t}^2 \cip 1$ and (b) $\max_{1 \leq t \leq T} |X_{T,t}| \cip 0$. 

       To show (a), note that by~\ref{ass:base}(i), the VAR admits the causal representation
    \[
    y_t=\sum_{j=0}^\infty \Theta_j\varepsilon_{t-j},
    \]
    with $\sum_{j=0}^\infty \|\Theta_j\|<\infty$, hence both  both $(u_t)_{t\in\Z}$ and $(V_t^2)_{t\in\Z}$ are stationary and ergodic. For some $\delta>0$, by Minkowski's inequality,
\[
\left(\E\|y_t\|^{2+\delta}\right)^{1/(2+\delta)}
\leq \sum_{j=0}^\infty \|\Theta_j\|
\left(\E\|\varepsilon_t\|^{2+\delta}\right)^{1/(2+\delta)}
<\infty,
\]
by~\ref{ass:beta}(iv). Moreover, since each $g_k(\varepsilon_t)$ is a polynomial of total degree of at most $r^*$, we have 
$\E\|g(\varepsilon_t)\|^{2 + \delta} < \infty$. Hence, $\E\|V_t\|^{2+\delta} \leq \E\|U_t\|^{2+\delta}  < \infty$. Hence, $(V_t^2)_{t \in \Z}$ is a stationary, ergodic and integrable sequence, by an ergodic Theorem,
    \[
    \sum^T_{t=1} X_{T,t}^2 = \frac1T \sum^T_{t=1} V_{t}^2/s_\lambda^2 \cip 1.
    \]
    To verify (b), it suffices to show that $\frac{1}{\sqrt{T}}\max_{1 \leq t \leq T} |V_t| \cip 0$.
    By monotonicity of the probability measure and Markov's inequality, we have for all $\epsilon>0$
    \begin{align*}
        \pr \left(\frac{1}{\sqrt{T}} \max_{1 \leq t \leq T} |V_t| > \epsilon \right) \leq \sum^T_{t=1} \pr \left(|V_t| > \epsilon \sqrt{T} \right) \leq \frac{1}{\epsilon^{2r} T^{r-1}}  \frac1T \sum^T_{t=1} \E |V_t|^{2r}.
    \end{align*}
    We have shown $\E |V_t|^{2r} < \infty$ for $r = 1 + \frac{\delta}{2} > 1$, the term is of order $O(T^{-(r-1)})$. Thus, we conclude $\frac{1}{\sqrt{T}}\sum_{t=1}^T V_t \ \Rightarrow \ \mc N (0, s_\lambda^2)$ and hence 
    $\frac1{\sqrt T}\sum_{t=1}^T U_t
\ \Rightarrow \ \mc N(0,S_{uu}).$
    
    To show that the moment functions in the first and second-step estimation have zero asymptotic covariance, note that 
    \[
S_{mg}
=
\E[(Z_{t-1}\otimes I_n)u_t g_t']
=
(\E[Z_{t-1}]\otimes I_n)\,B_0\,\E[\varepsilon_t g(\varepsilon_t)'].
\]
Hence $S_{mg}=\bs 0$ whenever $\E[Z_{t-1}]=0$ or $\E[\varepsilon_t g(\varepsilon_t)']= \bs 0$. In particular, if the process $(y_t)_t$ has zero mean and the VAR model contains no intercept, then
\[
\E[Z_{t-1}] = 
	\begin{pmatrix}
		\sum_{j=0}^\infty \Theta_{j} \E[\varepsilon_{t-j-1}]\\
		\vdots \\
		 \sum_{j=0}^\infty \Theta_{j}  \E[\varepsilon_{t-j-p}]
	\end{pmatrix}	= \bs 0_{np \times 1},
\]
and we have $S_{mg}=\bs 0$. This completes the proof.
\end{proof}

% -------------------------------------------- &
% -------------------------------------------- &

\begin{lemma} \label{lem:LLN_ULLN}
Under Assumptions of Proposition~\ref{thm:asymptotics}, let
\[
G_{\alpha,t}(\theta):=\frac{\partial g(y_t,\theta)}{\partial \alpha'}
\in \R^{K\times n^2p},
\qquad
G_{\beta,t}(\theta):=\frac{\partial g(y_t,\theta)}{\partial \beta'}
\in \R^{K\times n^2},
\]
and define $M := \Gamma \otimes I_n$. Let $\hat M_T = \left(\frac{ZZ'}{T}\right) \otimes I_n$, then
\begin{enumerate}[label=(\roman*)]
    \item $\hat M_T \cip M$;
    \item 
    \[
    \sup_{\theta\in\bar{\mc{N}}}
    \left\|
    \frac1T\sum_{t=1}^T G_{\alpha,t}(\theta)-\E[G_{\alpha,t}(\theta)]
    \right\|
    \cip 0,
    \qquad
    \sup_{\theta\in\bar{\mc{N}}}
    \left\|
    \frac1T\sum_{t=1}^T G_{\beta,t}(\theta)-\E[G_{\beta,t}(\theta)]
    \right\|
    \cip 0,
    \]
    where $\bar{\mc{N}} \subset \mc{N}$ is a compact subset of $\mc{N}$ containing $\theta_0$.
\end{enumerate}
\end{lemma}
\begin{proof}[Proof of Lemma~\ref{lem:LLN_ULLN}]
	We can write
    \[
Z_{t-1}= 
	\begin{pmatrix}
		\sum_{j=0}^\infty \Theta_{j} \varepsilon_{t-j-1}\\
		\vdots \\
		 \sum_{j=0}^\infty \Theta_{j}  \varepsilon_{t-j-p}
	\end{pmatrix}
    =	\begin{pmatrix}
		\sum_{j=1}^\infty \Theta_{j-1} \varepsilon_{t-j}  \\
		\vdots \\
		\sum_{j=1}^\infty \Theta_{j-p} \varepsilon_{t-j} 
	\end{pmatrix} 
    = \sum_{j=1}^\infty C_j \varepsilon_{t-j},
\]
for  $C_j \in \R^{np\times n}$, and
	\begin{align*}
	   \hat \Gamma_T
    = \frac1T\sum_{t=1}^T Z_{t-1}Z_{t-1}' = 
    \frac{1}{T}\sum_{t=1}^T \sum_{j_1=1}^\infty \sum_{j_2=1}^\infty C_{j_1} \varepsilon_{t-j_1} \varepsilon_{t-j_2}' C_{j_2} '
\end{align*}
By stability,  $\sum_{j=1}^\infty \|C_j\|<\infty$. Since both $( Z_t)_t$, and $( Z_t Z_t')_t$ are stationary and ergodic by Thm.\ 14.9 in \cite{Davidson2021}, and  by Minkowski's inequality and stationarity,
\[
\left(\E\|Z_{t-1}\|^2\right)^{1/2}
=
\left(
\E\left\|\sum_{j=1}^\infty C_j \varepsilon_{t-j}\right\|^2
\right)^{1/2}
\le
\bigl(\E\|\varepsilon_{t}\|^2\bigr)^{1/2} \sum_{j=1}^\infty \|C_j\| 
<\infty,
\] 
since  $\E[\varepsilon_t\varepsilon_t']=I_n$, $\E\|Z_{t-1}\|<\infty$ and
$\E\|Z_{t-1}Z_{t-1}'\|<\infty$.
The ergodic Theorem entails that $\frac1T\sum_{t=1}^T Z_{t-1}Z_{t-1}'
\to\E[Z_{t-1}Z_{t-1}'] = \Gamma$ a.s. and hence, in probability. The conclusion follows by continuity of the Kronecker product. 

To prove (ii), for $\theta \in \bar{\mc{N}}$ denote the vectorized Jacobian $q_t^\bullet(\theta) := \text{vec}(G_{\bullet,t}(\theta))$ 
and define $Q_T^\bullet(\theta) := \frac1T\sum_{t=1}^T q_t^\bullet(\theta)-\E[q_t^\bullet(\theta)]$ for $\bullet \in \{\alpha, \beta\}$.
Since $e_t(y_t,\theta)$ depends on the finite block $(y_t,y_{t-1},\dots,y_{t-p})$, and \(G_{\alpha,t}(\theta)\), \(G_{\beta,t}(\theta)\) are measurable (continuous) functions of $e_t(y_t,\theta)$, by Theorem~14.9 in \cite{Davidson2021}, $(q_t^\alpha(\theta))_t$ and $(q_t^\beta(\theta))_{t}$ are stationary and ergodic. By \ref{ass:beta}(iii), for all $\theta \in \bar{\mc{N}} \subset \mc{N}$, $\E\|q_t^\bullet(\theta)\| <\infty$ and the ergodic theorem entails that 
$Q_T^\bullet(\theta) \to 0$ a.s.\ as $T \to \infty$ pointwise on $\bar{\mc{N}}$ for both $\bullet \in \{\alpha, \beta\}$.

Next we show that $\{Q_T^\bullet(\theta)\}$ is asymptotically uniformly stochastically equicontinuous.
For each $1\leq t \leq T$,  $\bullet \in \{\alpha, \beta\}$, define 
\[
\Delta_t^\bullet(\delta)
:=\sup_{\theta \in \bar{\mc{N}}}
\sup_{\substack{\theta'\in \bar{\mc{N}} \\ \|\theta-\theta'\|<\delta}}
\|q_t^\bullet(\theta)-q_t^\bullet(\theta')\|,
\]
since $\theta\mapsto g(y_t,\theta)$ is continuously differentiable a.s.\ on $\mc N$, 
$
\lim_{\delta \downarrow 0} \Delta_t^\bullet(\delta,\omega) = 0
$
a.s. Since $\Delta_t^\bullet(\delta)
\leq 2\sup_{\theta\in \bar{\mc{N}}}\|q_t^\bullet(\theta)\|,$ and $\E\sup_{\theta\in\bar{\mc{N}}}\|q_t^\bullet(\theta)\| < \infty$ by \ref{ass:beta}(iii),
the dominated convergence theorem entails that $\E[\Delta_t^\bullet(\delta)]\to 0 $ as $\delta\downarrow 0$ for both $\bullet \in \{\alpha, \beta\}$.

Define the (local) modulus of continuity for both $\bullet \in \{\alpha, \beta\}$:
\begin{align*}
    w(Q_T^\bullet,\delta)
&:=
\sup_{\theta \in \bar{\mc{N}}}
\sup_{\substack{\theta'\in \bar{\mc{N}} \\ \|\theta-\theta'\|<\delta}}
\|Q_T^\bullet(\theta)-Q_T^\bullet(\theta')\| \\
&=\sup_{\theta \in \bar{\mc{N}}}
\sup_{\substack{\theta'\in \bar{\mc{N}} \\ \|\theta-\theta'\|<\delta}}
\left \|\frac1T\sum_{t=1}^T q_t^\bullet(\theta)-\E[q_t^\bullet(\theta)] - \frac1T\sum_{t=1}^T q_t^\bullet(\theta')+\E[q_t^\bullet(\theta')]  \right\|\\
&\leq \frac1T\sum_{t=1}^T \Delta_t^\bullet(\delta)+\E[\Delta_t^\bullet(\delta)].
\end{align*}
Hence, for any $\epsilon > 0$,
\[
\{w(Q_T^\bullet,\delta)\} \subseteq \left\{ \frac1T\sum_{t=1}^T \Delta_t^\bullet(\delta)+\E[\Delta_t^\bullet(\delta)] > \epsilon \right\} 
\subseteq \left\{\frac1T\sum_{t=1}^T \Delta_t^\bullet(\delta) > \epsilon/2\right\} \cup \{\E[\Delta_t^\bullet(\delta)] > \epsilon/2\},
\]
where the last inclusion can be seen by noting the contrapositive, i.e., 
\[
\left\{\frac1T\sum_{t=1}^T \Delta_t^\bullet(\delta) 
\leq \epsilon/2 \right\} \cap \{\E[\Delta_t^\bullet(\delta)] \leq \epsilon/2\} \subseteq \left\{\frac1T\sum_{t=1}^T \Delta_t^\bullet(\delta)+\E[\Delta_t^\bullet(\delta)] \leq \epsilon \right\}.
\]
Hence, by monotonicity of the probability measure, we have that for every $\epsilon > 0$
\[
\pr\left(w(Q_T^\bullet,\delta)>\epsilon\right)
\le
\pr\left(\frac1T\sum_{t=1}^T \Delta_t^\bullet(\delta)>\frac{\epsilon}{2}\right)
+
\mathbf 1\left\{\E[\Delta_t^\bullet(\delta)]>\frac{\epsilon}{2}\right\}.
\]
By Markov's inequality and due to stationarity,
\[
\limsup_{T\to\infty}\pr\left(w(Q_T^\bullet,\delta)>\epsilon\right)
\le
\frac{2}{\epsilon}\,\E[\Delta_t^\bullet(\delta)]
+
\mathbf 1\!\left\{\E[\Delta_t^\bullet(\delta)]>\frac{\epsilon}{2}\right\}.
\]
Letting $\delta\downarrow 0$ yields
\[
\lim_{\delta\downarrow0}\,
\limsup_{T\to\infty}\pr\left(w(Q_T^\bullet,\delta)>\epsilon\right)=0,
\]
and thus $\{Q_T^\bullet(\theta)\}$ is asymptotically uniformly stochastically equicontinuous. Therefore, by Thm.\ 22.9 of \cite{Davidson2021}, we obtain uniform convergence of the sample Jacobian.
\end{proof}

\begin{lemma}
\label{lem:scale_noise}
Let $g_t(\theta)=(g_1(y_t,\theta),\ldots,g_K(y_t,\theta))'$ be given by \eqref{eq: f}. Then, for every $\theta=(\alpha_0',\beta')'$ and every positive diagonal matrix $D=\diag(d_1,\ldots,d_n)$, the scaled moment vector satisfies
\begin{equation}
\label{eq:g_scaled_vector}
g_t(\theta_D)
=
\Delta(D)\,g_t(\theta)
+
\bigl(\Delta(D)-I_K\bigr)\bs c .
\end{equation}
Then, the noise term is equal to
\begin{align}
\label{lem_eq:trace_scaled_general}
\frac{1}{T}\trace\bigl(W S_{gg}(\theta_D)\bigr)
&=
\frac{1}{T} \sum_{k=1}^K \sum_{j=1}^K \delta_k \delta_j W_{kj} S_{gg, kj}(\theta) 
\\ \nonumber
&\quad
+
\frac{2}{T}\,
\sum_{k=1}^K \sum_{j=1}^K c(k) (\delta_k - 1) \delta_j W_{kj} \E[g_j(y_t, \theta)] 
\\ \nonumber
&\quad
+
\frac{1}{T}\,
\sum_{k=1}^K \sum_{j=1}^K c(k)  c(j)  (\delta_k - 1) (\delta_j - 1) W_{kj}.
\end{align}
% \begin{align}
% \label{eq:Sgg_scaled}
% S_{gg}(\theta_D)
% &=
% \Delta(D)\,S_{gg}(\theta)\,\Delta(D)
% +
% \Delta(D)\,\mu_g(\theta)\,c'\bigl(\Delta(D)-I_K\bigr)
% \\ \nonumber
% &\quad
% +
% \bigl(\Delta(D)-I_K\bigr)c\,\mu_g(\theta)'\,\Delta(D)
% +
% \bigl(\Delta(D)-I_K\bigr)cc'\bigl(\Delta(D)-I_K\bigr).
% \end{align}

\end{lemma}

\begin{proof}[Proof of Lemma~\ref{lem:scale_noise}]
Fix the $k$-th moment condition in \eqref{eq: f}. Since $e_t(\theta_D)=D^{-1}e_t(\theta)$, we have
\begin{align*}
g_k(y_t,\theta_D)
&=
\prod_{i=1}^n e(\theta_D)^{k_i}_{i,t}-c(k)
=
\prod_{i=1}^n \bigl(d_i^{-1}e(\theta)_{i,t}\bigr)^{k_i}-c(k)
\\
&=
\delta_k(D)\prod_{i=1}^n e(\theta)_{i,t}^{k_i}-c(k)
\\
&=
\delta_k(D)\Bigl(\prod_{i=1}^n e(\theta)_{i,t}^{k_i}-c(k)\Bigr)
+\bigl(\delta_k(D)-1\bigr)c(k)
\\
&=
\delta_k(D)\,g_k(y_t,\theta)+\bigl(\delta_k(D)-1\bigr)c(k).
\end{align*}
Stacking the $K$ moment conditions yields \eqref{eq:g_scaled_vector}.
Next, plugging in \eqref{eq:g_scaled_vector} yields
\begin{align*}
S_{gg}(\theta_D) 
&= \E\!\left[g_t(\theta_D)g_t(\theta_D)'\right]
=
\E\!\left[
\bigl(\Delta(D) g_t(\theta)+(\Delta(D)-I_K)\bs c\bigr)
\bigl(\Delta(D) g_t(\theta)+(\Delta(D)-I_K) \bs c\bigr)'
\right]
\\
&=
\Delta(D)\E[g_t(\theta)g_t(\theta)']\Delta(D)
+
\Delta(D) \E[g_t(\theta)]\,\bs c'(\Delta(D)-I_K)
\\
&\quad
+
(\Delta(D)-I_K)\bs c\,\E[g_t(\theta)]'\Delta(D)
+
(\Delta(D)-I_K)\bs c \bs c'(\Delta(D)-I_K),
\end{align*}
Finally,
\begin{align*}
\trace\bigl(W S_{gg}(\theta_D)\bigr)
&=
\trace\bigl(W\Delta(D)S_{gg}(\theta)\Delta(D)\bigr)
+
2 \bs c'\bigl(\Delta(D)-I_K\bigr)W\Delta(D)\E[g_t(\theta)]
\\ 
&\quad
+
\bs c'\bigl(\Delta(D)-I_K\bigr)W\bigl(\Delta(D)-I_K\bigr) \bs c\\
&= \sum_{k=1}^K \sum_{j=1}^K \delta_k \delta_j W_{kj} S_{gg, kj}(\theta) 
+ 2 \sum_{k=1}^K \sum_{j=1}^K c(k) (\delta_k - 1) \delta_j W_{kj} \E[g_j(y_t, \theta)] \\
&\quad
+\sum_{k=1}^K \sum_{j=1}^K c(k)  c(j)  (\delta_k - 1) (\delta_j - 1) W_{kj},
\end{align*}
where the two cross terms coincide by symmetry of $W$.
\end{proof}

% \clearpage
\section{GMM under pre-whitening}
\label{app:prewhitening}
This section compares the GMM estimator introduced in Section~\ref{sec:svar_gmm} with an estimator that pre-whitens the reduced-form residual in the first step. Unlike the proposed approach where we estimate only the reduced-form slope parameter $\alpha$, a pre-whitened ICA-type procedure also estimates the reduced-form covariance matrix $\Sigma_u :=  \E[u_t u_t'] = B_0 B_0'$ in the first step. This will make the covariance condition $\E[\varepsilon_t \varepsilon_t‘] = I_n$ binding and thereby free from the scale-direction bias in small sample discussed in Section~\ref{sec: bias}. However, as we illustrate below, the additional first-step scale estimation destroys the adaptivity property discussed in Corollary~\ref{corol:G_a}, such that the weighting $S_{gg}^{-1}$ is no longer optimal. The optimal weighting in this case requires taking the estimation uncertainty of reduced-form scale parameters into account.

Let $\omega$ be a smooth parametrization of the positive definite reduced-form covariance matrix $\Sigma_u$. Without loss of generality (w.l.o.g.) and to simplify the notation, let $\omega := \text{vec}(\Sigma_u)$ and write the contemporaneous map between the structural and reduced-form errors as
\[
y_t - (Z_{t-1}' \otimes I_n)\alpha = u_t = \Sigma_u^{1/2} Q_0 \varepsilon_t,
\]
where $\Sigma_u^{1/2} $ is any matrix square-root of $\Sigma_u$ uniquely identified via the Cholesky or spectral decomposition, and $Q_0$ is orthogonal such that $Q_0 Q_0'= Q_0'Q_0 = I_n$. W.l.o.g., suppose $\Sigma_u^{1/2}$ is the symmetric square root such that $\Sigma_u = \Sigma_u^{1/2} \Sigma_u^{1/2}$.
Note that $B_0 = \Sigma_u^{1/2} Q_0$. Let $Q$ be parametrized by an $n(n-1)/2$-dimensional structural parameter vector $\rho \in \R^{n(n-1)/2}$.\footnote{Note that $Q_0 Q_0'= Q_0'Q_0 = I_n$ imposes $n(n+1)/2$ conditions so that there remain only $n(n-1)/2$ free parameters. 
A useful way of parameterizing matrix $Q$ is to formulate it as the product of a sequence of $n(n-1)/2$ Givens rotation matrices. For example, in the case of $n=3$, we have
	\begin{align*}
	Q(\rho) =
	\begin{bmatrix}
		1 & 0 & 0 \\
		0 & \cos \rho_1 & -\sin \rho_1 \\
		0 & \sin \rho_1 & \cos \rho_1 
	\end{bmatrix}
	\begin{bmatrix}
	 \cos \rho_2 & 0 & -\sin \rho_2 \\
	 0 & 1 & 0 \\
	 \sin \rho_2 & 0 & \cos \rho_2
	\end{bmatrix}
	\begin{bmatrix}
	 \cos \rho_3 & -\sin \rho_3 & 0 \\
	 \sin \rho_3 & \cos \rho_3 & 0 \\
	 0 & 0 & 1
	\end{bmatrix}.
\end{align*}
See \cite{gourieroux2017statistical, herwartz2023srhl, hafner2025kml, Wang2026QE} for details.}
Given the reduced-form parameters $(\alpha, \omega)$, define the pre-whitened residuals 
$z_t(\alpha,\omega) := \Sigma_u^{-1/2}(\omega)u_t(\alpha)$ and the moment functions on the structural innovations in the second step 
are applied to 
\[
e_t(\alpha,\omega,\rho) := Q(\rho)' z_t(\alpha,\omega) = Q(\rho)' \Sigma_u^{-1/2}(\omega) u_t(\alpha).
\]
At the true value $(\alpha_0,\omega_0,\rho_0)$, we have $Q(\rho_0) = Q_0$, $\Sigma_u^{-1/2}(\omega_0) = \Sigma_u^{-1/2}$, and hence $e_t(\alpha_0,\omega_0,\rho_0) = Q_0' \Sigma_u^{-1/2} \Sigma_u^{1/2} Q_0 \varepsilon_t = \varepsilon_t$. 

Let $s_t(\alpha, \omega) \coloneq \text{vec} \left(u_t(\alpha) \, u_t(\alpha)' - \Sigma_u(\omega) \right)$. Given the LS estimator for the VAR slope parameters $\hat\alpha_T$, let $\hat\omega_T$ be an estimator that solves $\bar s_T(\hat \alpha_T, \omega) \coloneqq \frac1T \sum^T_{t=1} s_t(\hat \alpha_T, \omega) = \bs 0$. Apart from the covariance conditions, let the higher moment conditions be based on the same map $g(\cdot)$ as in Section~\ref{sec:svar_gmm}, the pre-whitened GMM estimator for the structural parameter $\rho$ is
\[
\hat\rho_T
:=
\argmin_{\rho}
\bar g_T(\hat\alpha_T,\hat\omega_T,\rho)'
\hat W_T
\bar g_T(\hat\alpha_T,\hat\omega_T,\rho).
\]
Denote the Jacobian blocks
\[
G_\alpha^* := \E\!\left[
\frac{\partial g_t(\alpha_0,\omega_0,\rho_0)}{\partial \alpha'}
\right],
\quad
G_\omega^*
:=
\E\!\left[
\frac{\partial g_t(\alpha_0,\omega_0,\rho_0)}{\partial \omega'}
\right],
\quad
G_\rho^*
:=
\E\!\left[
\frac{\partial g_t(\alpha_0,\omega_0,\rho_0)}{\partial \rho'}
\right].
\]
It can easily be shown that under conditions similar to those in Assumption~\ref{ass:beta}, the estimator $\hat\rho_T$ satisfies\footnote{
Note that $G_\alpha^* = -\E[\dot g_\varepsilon(\varepsilon_t) Q_0' \Sigma_u^{-1/2} (Z_{t-1}'\otimes I_n) ] = -\E[\dot g_\varepsilon(\varepsilon_t) B_0^{-1}(Z_{t-1}' \otimes I_n)]$ and by the same argument used in Corollary~\ref{corol:G_a}, we have $G_\alpha^* = \bs 0$. For $\hat\omega_T$, denote $S_\alpha \coloneqq \E\left[\frac{\partial s_t(\alpha_0, \omega_0)}{\partial \alpha'} \right]$ and $S_\omega \coloneqq  \E\left[\frac{\partial s_t(\alpha_0, \omega_0)}{\partial \omega'} \right].$ By our parametrization, we have $S_\omega = -I_{n^2}$, while $\E[u_t(\alpha_0)Z_{t-1}']=0$ and therefore $S_\alpha = - \E\left[((u_t(\alpha_0) \otimes I_n) + ( I_n \otimes u_t(\alpha_0) ) )(Z_{t-1}' \otimes I_n)\right] = \bs 0$. Expanding the moment function around the true value and inserting the influence function for $\hat \alpha_T$ yields
 \begin{align*}
        \sqrt{T}(\hat \omega_T - \omega_0) &= \frac{1}{\sqrt{T}}\sum_{t=1}^T -S_\omega^{-1} \left( s_t(\alpha_0, \omega_0) - S_\alpha M^{-1} m_t(\alpha_0) \right) + o_p(1)\\
        &= \frac{1}{\sqrt{T}}\sum_{t=1}^T s_t(\alpha_0, \omega_0) + o_p(1).
\end{align*}
}
\[
\sqrt{T}(\hat\rho_T-\rho_0) =
-(G_\rho^{*\prime}WG_\rho^*)^{-1}G_\rho^{*\prime}W
\frac1{\sqrt{T}}
\sum_{t=1}^T
\left( g_t(\alpha_0,\omega_0,\rho_0) + G_\omega^* s_{t}(\alpha_0, \omega_0) \right) + o_p(1),
\]
where
\[
G_\omega^* = \E\!\left[
\dot g_\varepsilon(\varepsilon_t)\,
Q_0' (\varepsilon_t' Q_0' \Sigma_u^{1/2}\otimes I_n) \frac{\partial \text{vec}(\Sigma_u^{-1/2})}{\partial \omega'}
\right],
\]
and $\frac{\partial \text{vec}(\Sigma_u^{-1/2})}{\partial \omega'}$ can be obtained by differentiating the identity 
$\Sigma_u^{-1/2} \Sigma_u \, \Sigma_u^{-1/2} = I_n$:
    \[
    \left[(\Sigma_u^{1/2} \otimes I_n) + (I_n \otimes \Sigma_u^{1/2}) \right] \mathrm{d} \text{vec}(\Sigma_u^{-1/2}) = -( \Sigma_u^{-1/2} \otimes \Sigma_u^{-1/2}) \mathrm{d} \text{vec}(\Sigma_u) =  -( \Sigma_u^{-1/2} \otimes \Sigma_u^{-1/2}) \mathrm{d} \omega.
    \]
As a consequence, the weighting $S_{gg}^{-1}$ does not lead to asymptotic efficiency, while the optimal weighting $(S_{\perp}^{*})^{-1}$ must explicitly take the estimation uncertainty of the whitening parameters into account, where 
\[
S_{\perp}^{*} = \E\!\left[\left(g_t(\alpha_0,\omega_0,\rho_0) + G_\omega^* s_{t}(\alpha_0, \omega_0) \right)  \left(g_t(\alpha_0,\omega_0,\rho_0) + G_\omega^* s_{t}(\alpha_0, \omega_0) \right)'\right],
\]
and $S_{\perp}^{*} \neq S_{gg}$ unless $G_\omega^* \neq \bs 0$.    
Note that in addition to $\dot g_\varepsilon$, the term in the Kronecker product in $G_\omega^*$ also depends on the current shock, so that $G_\omega^* \neq \bs 0$ in general. 

To illustrate, consider for $n=2$ a co-skewness condition $\E[\varepsilon_{1,t}^2 \varepsilon_{2,t}] = 0$, so that 
$
\dot g_\varepsilon(\varepsilon_t)'= 
\begin{pmatrix}
2\varepsilon_{1,t}\varepsilon_{2,t}\\
\varepsilon_{1,t}^2
\end{pmatrix}.
$ 
Let the structural contemporaneous map be parameterized as
\[
u_t = \Sigma_u^{1/2}(\omega) Q(\rho) \varepsilon_t
=
\begin{pmatrix}
\omega_1 & 0\\
\omega_2 & \omega_3
\end{pmatrix}
\begin{pmatrix}
\cos\rho & -\sin\rho\\
\sin\rho & \cos\rho
\end{pmatrix}
\begin{pmatrix}
\varepsilon_{1,t}\\
\varepsilon_{2,t}
\end{pmatrix},
\]
with $\omega_1 > 0, \ \omega_3 > 0$. W.l.o.g., suppose $\rho_0 =0$ such that $Q_0 = I_2$, then, 
\[
\frac{\partial}{\partial \omega_2}e_t(\omega_0,0) 
= \left[ \frac{\partial}{\partial \omega_2}
\begin{pmatrix}
\omega_{1,0}^{-1} & 0\\[0.3em]
-\dfrac{\omega_{2,0}}{\omega_{1,0} \omega_{3,0}} & \omega_{3,0}^{-1}
\end{pmatrix}\right] \Sigma_u^{1/2} \varepsilon_t = 
\begin{pmatrix}
0\\ 
-\dfrac{1}{\omega_{3,0}}\varepsilon_{1,t}
\end{pmatrix}.
\]
and the expected partial derivative with respect to $\omega_2$ at the true value is
\[
\frac{\partial}{\partial \omega_2}
\E[g(e_t(\omega_0,0))]
=
\E\!\left[
\dot g_\varepsilon(\varepsilon_t)
\frac{\partial}{\partial \omega_2}e_t(\omega_0,0)
\right]
=
-\frac{1}{\omega_{3,0}}  \E[\varepsilon_{1,t}^3].
\]
Therefore, the expected Jacobian with respect to the whitening parameter $\omega_2$ is zero iff the first structural shock has zero skewness, i.e., $\E[\varepsilon_{1,t}^3] = 0$.

% \clearpage
\section{Structured estimation of the expected Jacobian}
\label{app:estimate_jacobian}

The assumption of mutually independent structural shocks can also be exploited to estimate the Jacobian  matrix $G_\beta$ required to estimate the asymptotic variance of the GMM estimator.
For an arbitrary moment condition 
$ g_k(y_t, \theta) = \prod_{i=1}^{n} e(\theta)_{i,t}^{k_i} - c(k) $ with $k \in  \mathbf{2} \cup \mathbf{3} \cup \mathbf{4}$
we can compute the entry-wise derivative with respect to $b_{pq}$, the element at row $p$ and column $q$ of $B$, evaluated at $B=B_0$ corresponds to an element of $G_\beta$ as
\begin{align}
G_{\beta, k,b_{pq}}&:= \E \! \left[ \frac{\partial  g_k(y_t, \theta_0) }{\partial b_{pq}} \right]
\\ 
\label{eq: G unc}
&= 
\sum_{j=1,j\neq q}^{n} -k_j  a_{j p} \E \! \left[ \varepsilon_{j,t}^{k_j-1} \varepsilon_{q,t}^{k_q+1} \prod_{i=1,i\neq j,q}^{n} \varepsilon_{i,t}^{k_i} \right] 
-
k_q a_{q p} \E \!\left[\prod_{i=1 }^{n} \varepsilon_{i,t}^{k_i}  \right] ,
\end{align} 
with $A=B_0^{-1}$ and $a_{jp}$ denoting the elements of $A$.  The equality follows from the product rule and $\frac{\partial  e(\theta_0)_{i,t} }{\partial b_{pq}} = -a_{ip} \varepsilon_{q,t}$. Using the assumption of mutually independent shocks allows us to decompose higher-order co-moments in Equation (\ref{eq: G unc}) into a product of lower-order moments such that
\begin{align}
\label{eq: G ind}
	G_{\beta, k ,b_{pq}}^{SMI} = 
 	\sum_{j=1,j\neq q}^{n} -k_j  a_{j p} 
	\E \! \left[ \varepsilon_{j,t}^{k_j-1} \right]
	 	\E \! \left[ \varepsilon_{q,t}^{k_q+1} \right]
	 	\prod_{i=1,i\neq j,q}^{n} 	\E \! \left[\varepsilon_{i,t}^{k_i} \right] 
	-
	k_q a_{q p} \prod_{i=1 }^{n} \E \! \left[ \varepsilon_{i,t}^{k_i}  \right] ,
\end{align}
where the superscript $SMI$ indicates that  the equality  $G_{\beta,m,b_{pq}} = G_{\beta,m,b_{pq}}^{SMI}$ holds for mutually independent shocks.

The assumption of serially and mutually independent shocks can also be used to   derive a   guess for the optimal weighting matrix $W^{*}$   without requiring an initial guess or estimate of the unknown simultaneous interaction $B_0$. Instead, the researcher can guess the distribution of each structural shock $\varepsilon_{i,t}$ for $i=1,...,n$ and if the guess is correct, Equation (\ref{eq: S SIMI})  directly yields the correct covariance matrix $S_{gg}$, which can be used to  calculate the optimal weighting matrix. 

% \clearpage

\section{Mean-independent covariance estimation}
\label{sec: appendix mi regularization}

Using the assumption of mean independent shocks to compute the covariance matrix $\widehat S^{SWI}$ factors moments into products of moments if possible due to the mean independence assumption, but retains joint moments otherwise. The resulting matrix $S^{SWI}$ converges to the true covariance matrix $S_{gg}$, however, in a finite sample, it need not be positive semidefinite. To stabilize inversion, we symmetrize it and write $\widehat S^{SWI}=DCD$, where $D$ contains its positive estimated moment standard deviations and $C$ has diagonal entries one. To ensure strict positivity, let $\lambda_{\min}$ and $\lambda_{\max}$ denote the extreme eigenvalues of $C$. With $c=100$, choose
\begin{equation}
 \begin{aligned}
 a&=\begin{cases}
 0,&\lambda_{\min}>0,\ \lambda_{\max}\leq c\lambda_{\min},\\
 \dfrac{\lambda_{\max}-c\lambda_{\min}}
 {\lambda_{\max}-c\lambda_{\min}+c-1},&\text{otherwise},
 \end{cases}
 \\[4pt]
 \widehat W_{\mathrm{SWI}}&=
 \{D[(1-a)C+aI]D\}^{-1}.
 \end{aligned}
 \label{eq:mc-mi-shrinkage}
\end{equation}
This procedure selects the smallest shrinkage intensity making the matrix positive definite with eigenvalue ratio at most 100. The transformation preserves the estimated diagonal variances. If the diagonal is unsuitable for correlation standardization, the implementation instead shrinks the covariance toward a positive scaled identity. The inversion uses a pseudoinverse when the reciprocal condition number is at most $10^{-12}$.
Note that in case the population correlation matrix is positive definite with eigenvalue ratio below 100, this repair algorithm becomes inactive asymptotically.

% \clearpage
\section{Additional Monte Carlo results}
\label{sec: appendix mc}

%\subsection{Additional results for the weighting comparison}
%\label{sec:mc-weighting-details}

\begin{figure}[H]
  \centering
  \includegraphics[width=\linewidth]{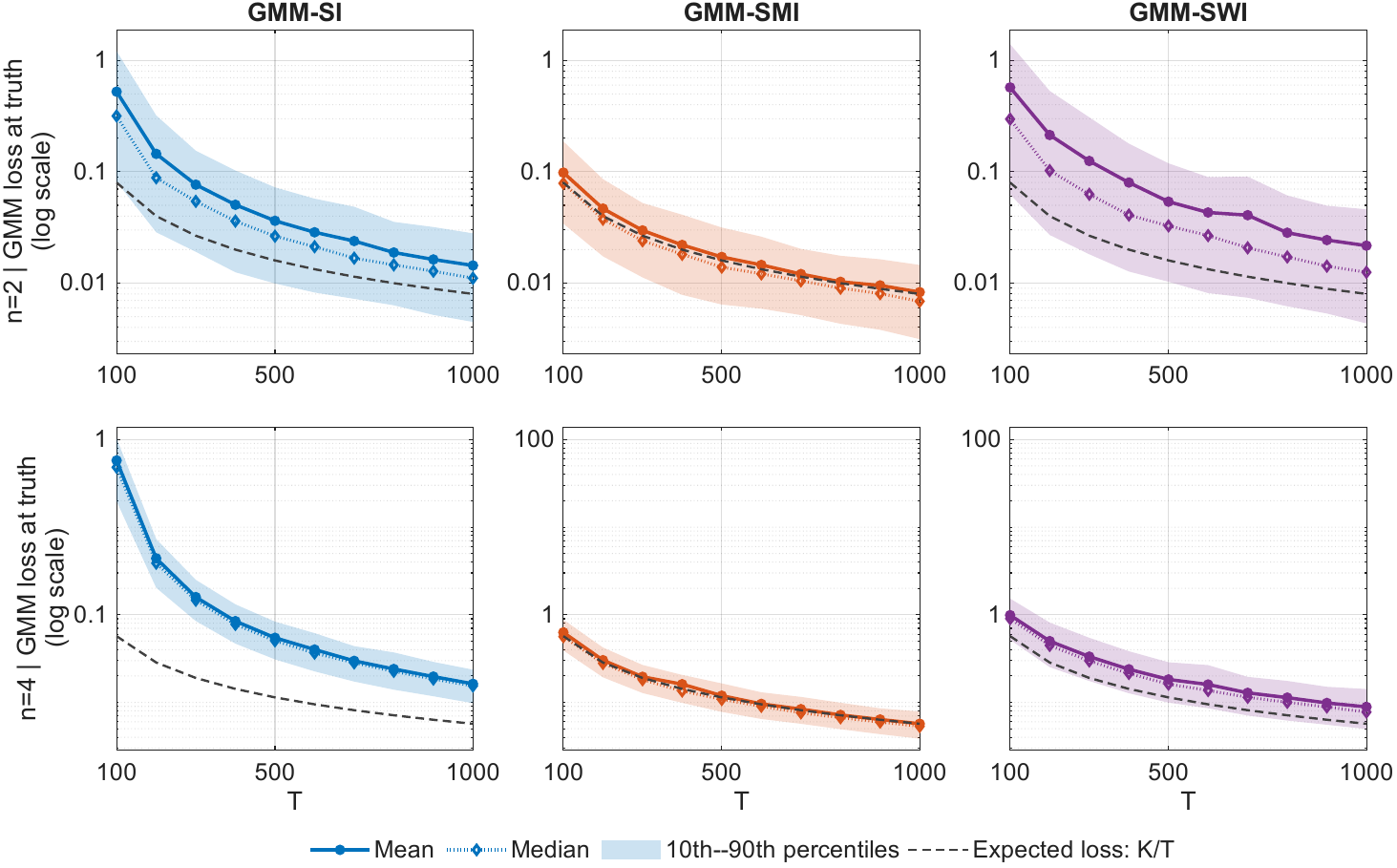}
  \caption{Log GMM loss under alternative weighting schemes.}
  \label{fig:mc-gmm-weighting-loss-full}
  \par\medskip
  \begin{minipage}{0.98\linewidth}
    \footnotesize
    \textit{Notes:} 
    Results use the design of Section~\ref{sec:mc-weighting}, with
    1,000 replications at each $T\in\{100,200,\ldots,1000\}$.
    Rows correspond to $n=2$ and $n=4$, while columns distinguish GMM-SI, GMM-SMI,
    GMM-SWI. Solid lines with circles show arithmetic means,
    dotted lines with diamonds show medians, and shaded regions span the
    10th--90th percentiles. Vertical axes are logarithmic. The dashed curve gives the expected loss by using the population optimal weight.
  \end{minipage}
\end{figure}

\begin{figure}[t]
  \centering
  \includegraphics[width=\linewidth]{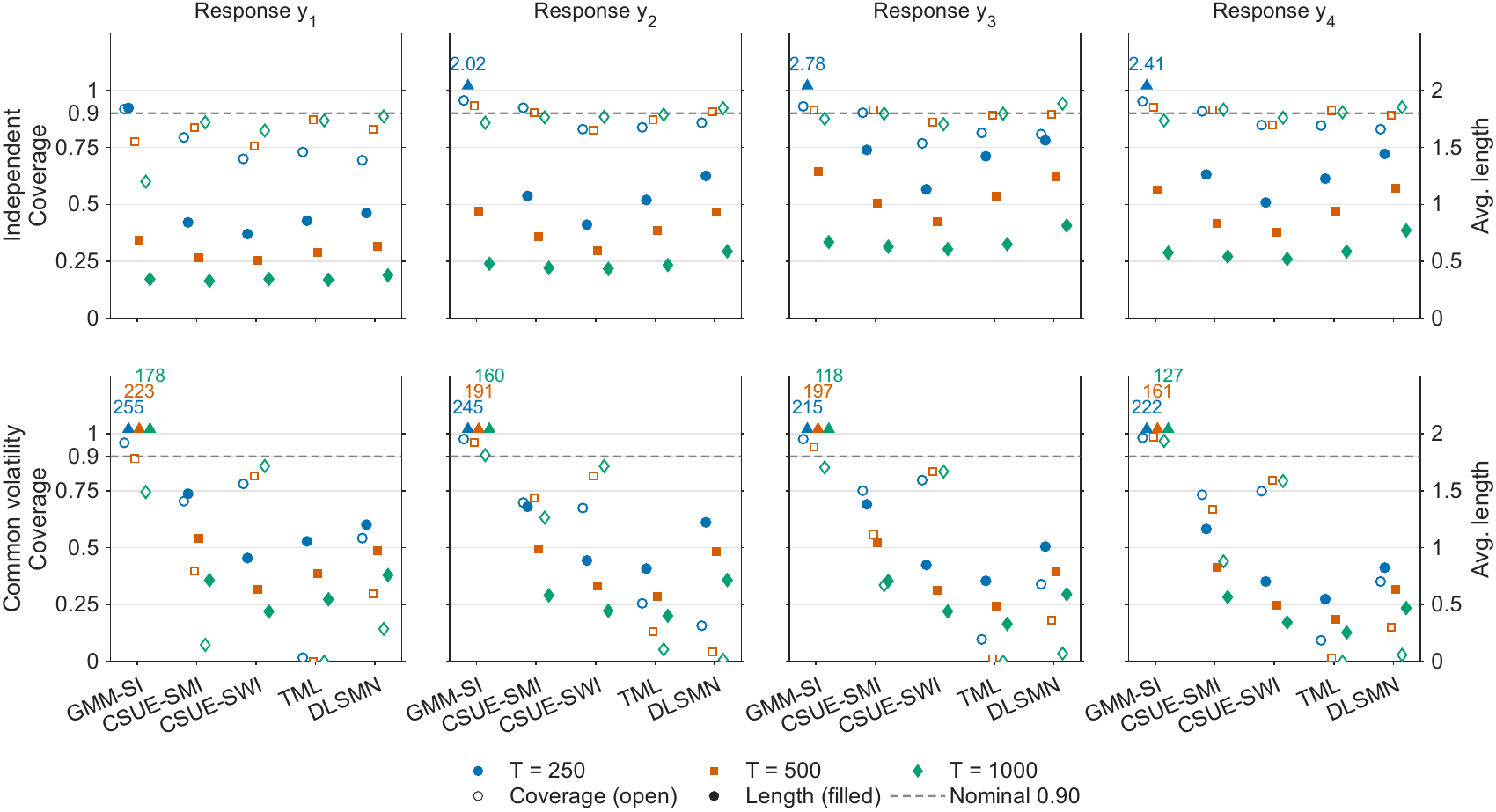}
  \caption{Coverage and average length of pointwise 90\% bootstrap intervals for impact responses.}
  \label{fig:mc-bootstrap-cv-dlsmn-saddle-shock1-h0}
  \par\medskip
  \begin{minipage}{0.98\linewidth}
    \footnotesize
    \textit{Notes:} 
    Open markers show coverage on the left axis; filled markers
    show average interval length on the right axis. Both axes are linear.
    The dashed line is the nominal coverage rate of 0.90. Rows, columns, and
    sample-size markers follow Figure~\ref{fig:mc-cv-dlsmn-saddle-shock1-h0-ranges-90}.
    Triangular callouts indicate lengths exceeding two, with the adjacent
    number giving the actual value. Results are based on 500 Monte Carlo replications with 500 bootstraps in each replication. 
  \end{minipage}
\end{figure}

 \subsection{Scale normalization}
\begin{figure}[t]
  \centering
  \includegraphics[width=\linewidth]{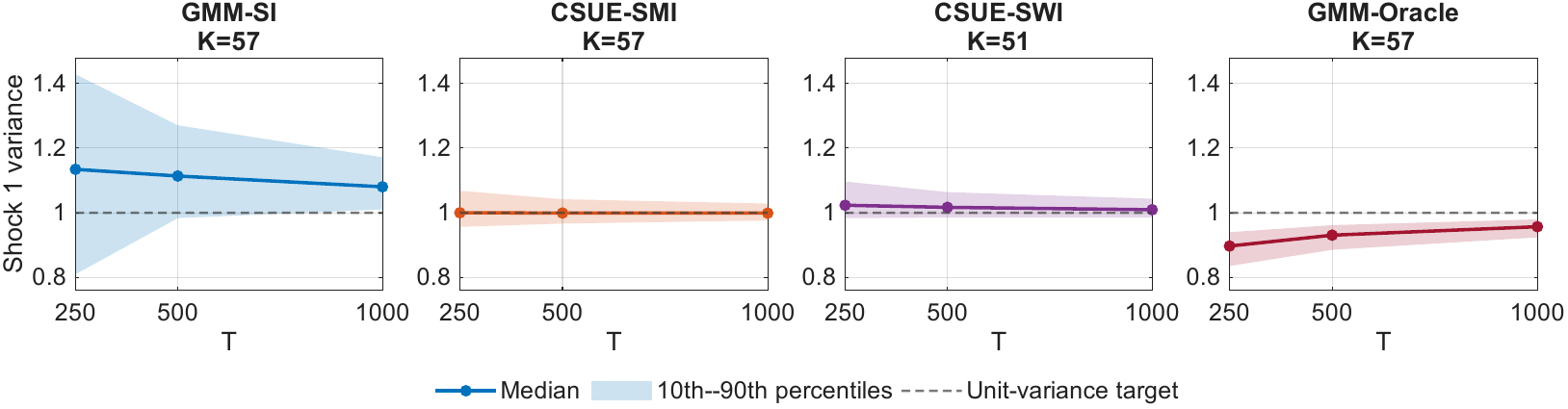}
  \caption{Finite-sample scaling of the estimated first structural shock.}
  \label{fig:mc-gmm-scaling-bias}
  \par\medskip
  \begin{minipage}{0.98\linewidth}
    \footnotesize
    \textit{Notes:} The figure reports the sample second moment
    $\widehat v_1=T^{-1}\sum_{t=1}^T\widehat\varepsilon_{1t}^{2}$ in the
    four-variable design across 1,000 replications for
    $T\in\{250,500,1000\}$. Lines with circles show Monte Carlo medians and
    shaded regions show the 10th--90th percentile ranges; the dashed line is
    the unit-variance target.  GMM-SI, CSUE-SMI,
and GMM-Oracle use $K=57$ moments implied by independent shocks and CSUE-SWI uses $K=51$ moments implied by mean independent shocks.
  \end{minipage}
\end{figure}

This simulation illustrates the scaling bias of GMM and how the continuous scale updating estimator addresses it. The simulation uses the GMM-SI, CSUE-SMI, and CSUE-SWI estimators from Section \ref{sec: Finite sample performance} and additionally shows the infeasible GMM-Oracle
using the population weight without a scale update.   The DGP is the four-dimensional SVAR(0) from Section \ref{sec:mc-weighting}. Figure \ref{fig:mc-gmm-scaling-bias} reports the sample second moment  $ \widehat v_1=\frac{1}{T}\sum_{t=1}^T
  \widehat\varepsilon_{1t}^{2}$
whose normalization target is one. At $T=250$, the median is $0.897$ for GMM-Oracle and $1.134$ for GMM-SI, compared with $1.000$ for CSUE-SMI and $1.023$ for CSUE-SWI. Even the oracle's 90th percentile, $0.940$, lies below one. At $T=1000$, the medians are $0.957$, $1.080$, $0.999$, and $1.009$, respectively. These results accord with the scaling distortion analyzed in Section~\ref{sec: bias} and show that the CSUE implementations substantially improve normalization.

\subsection{Distributional calibration and scope}
\label{sec:mc-dgp-details}

The impact matrices used in the Monte Carlo experiments are given by
\begin{equation}
  B_0^{(2)}=
  \begin{bmatrix}
    2.8 & 0 \\
   -0.6 & 2.2
  \end{bmatrix},
  \qquad
  B_0^{(4)}=
  \begin{bmatrix}
     2.8 &  0   & 0   & 0 \\
    -0.6 &  2.2 & 0   & 0 \\
    -1.6 &  1.1 & 2.6 & 0 \\
     1.2 & -0.9 & 0.7 & 2.4
  \end{bmatrix}.
  \label{eq:mc-gmm-focused-b0}.
\end{equation}
The independent-shock distribution used in both experiments is the standardized Pearson type-IV law with Pearson-system coefficients $(c_0,c_1,c_2)=(0.625,-0.25,0.125)$. For the common-volatility design in Section~\ref{sec:mc-comparison}, the base innovation follows
\begin{equation}
 z\sim 0.4\,\mathcal N(\mu_-,\sigma_-^2)
       +0.6\,\mathcal N(\mu_+,\sigma_+^2),
 \label{eq:mc-base-mixture}
\end{equation}
where $(\mu_-,\sigma_-,\mu_+,\sigma_+)$ equals
\[
 (-0.989728409296391,\;0.862152705707340,\;
   0.659818939530927,\;0.287621091516337).
\]
These values give $(\E[z],\E[z^2],\E[z^3],\E[z^4])=(0,1,-1,3.05)$ up to numerical rounding. The innovations are independent across dates and components and independent of all volatility paths. For $\pi_H=0.30$ and $m_2=10/3.05$, the variance states solve $\E[W]=1$ and $\E[W^2]=m_2$:
\begin{equation}
 w_L=1-\sqrt{(m_2-1)\frac{\pi_H}{1-\pi_H}},
 \qquad
 w_H=1+\sqrt{(m_2-1)\frac{1-\pi_H}{\pi_H}}.
\end{equation}
The implied standard-deviation multipliers are $0.108531$ and $1.818199$, approximately. With rows and columns ordered low, high, each path has transition matrix
\[
 P_R=\begin{bmatrix}0.775&0.225\\0.525&0.475\end{bmatrix}.
\]
Shocks 1 and 2 share their entire path, while each of shocks 3 and 4 has an independent path. Consequently,
\[
 \E[\varepsilon_i^3]=-\E[W^{3/2}]=-1.8041027,
 \quad \E[\varepsilon_i^4]=10,
 \quad \E[\varepsilon_1^2\varepsilon_2^2]=\E[W^2]=3.2786885.
\]

\subsection{Optimization, normalization, and bootstrap construction}
\label{sec:mc-implementation}

The point-estimate GMM procedures use an identity-weight preliminary estimate, initialized from the Cholesky factor of the residual second-moment matrix. They then construct the estimator-specific base weight at that preliminary estimate. GMM-SI keeps its second-step weight fixed. Both CSUE procedures update only the scale term during the final optimization. It is worth stressing that no zero restrictions on the impact matrix are imposed during any estimation procedure considered.

For point estimates, the sign and permutation are selected following the convention of setting diagonal entries to be positive and permuting the columns to maximize the trace. In each bootstrap replication, the estimates using the bootstrapped the data are matched by minimum Frobenius distance over signed permutations to the original estimate given the original data generated in that experiment. 

Bootstrap confidence intervals for the (scalar) structural impact multipliers are computed as follows. Each bootstrap draw samples a block (with length of eight) of residuals and reconstructs the data recursively \citep[see][for details and validity of the bootstrap for inference about structural IRFs]{BJT_MBB2016}. For an initial estimator $\widehat\theta$ and bootstrap quantiles $q_{.05}^*,q_{.50}^*,q_{.95}^*$, the pointwise 90\% confidence interval is computed as
\begin{equation}
 \left[\widehat\theta-|q_{.05}^*-q_{.50}^*|,\;
       \widehat\theta+|q_{.95}^*-q_{.50}^*|\right].
 \label{eq:mc-bootstrap-interval}
\end{equation}
The average lengths of the confidence intervals reported in Section~\ref{sec:mc-comparison} are computed across the 500 Monte Carlo replications.

% \clearpage
\section{Additional results for the oil-market application}
\label{app:application}
\begin{figure}[ht]
\centering
\includegraphics[width=0.5\linewidth]{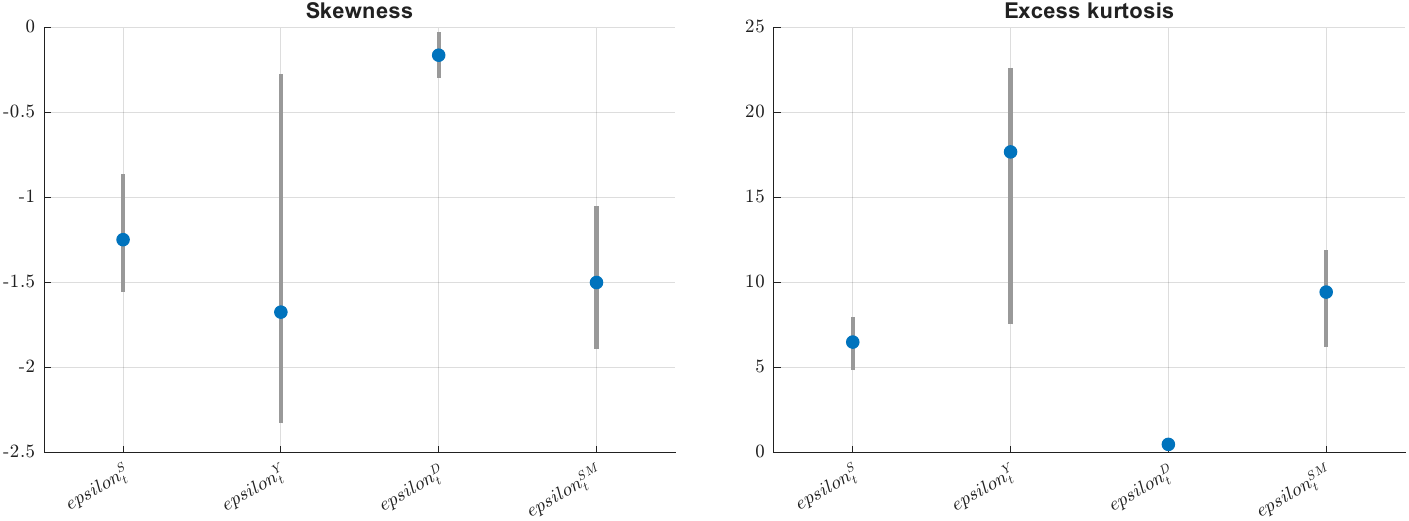}
\caption{Non-Gaussianity diagnostics for structural shocks from CSUE-SWI with $68$\% bootstrap confidence intervals.}
\label{fig:application_oilbridge_newdata_robust_epsilon_csue_meanindependent_nongaussianity}
\end{figure}
\begin{figure}[ht]
\centering
\includegraphics[width=0.5\linewidth]{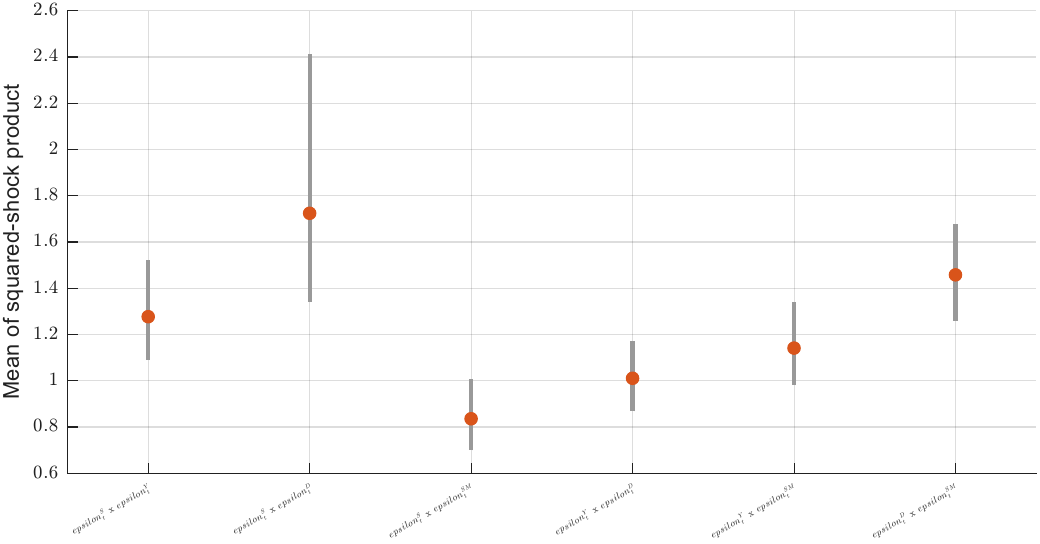}
\caption{Dependence in squared CSUE-SMI structural shocks. See Figure~\ref{fig:oil-dependence} for further details.}
\label{fig:application_oilbridge_newdata_robust_epsilon_csue_independent_fullmoments_dependence}
\end{figure}

This appendix supplements Section~\ref{sec:Application} with distributional diagnostics and a comparison of
CSUE under mean and mutual independence.

Figure \ref{fig:application_oilbridge_newdata_robust_epsilon_csue_meanindependent_nongaussianity} shows the skewness and excess kurtosis of the estimated structural shocks, indicating that three of the four shocks are non-Gaussian.
Figure \ref{fig:application_oilbridge_newdata_robust_epsilon_csue_independent_fullmoments_dependence} shows the common volatility of the shocks identified using the CSUE-SMI estimator. In comparison to the CSUE-SWI estimator in Section \ref{sec:Application}, the CSUE-SMI estimator additionally includes symmetric cokurtosis moment conditions that penalize deviations of $
\frac{1}{T}\sum_{t=1}^{T}
\widehat\varepsilon_{it}^{2}\widehat\varepsilon_{jt}^{2}
$ from $1$. The figure shows that the CSUE-SMI estimator is unable to satisfy these symmetric cokurtosis conditions. However, a comparison with Figure \ref{fig:oil-dependence} illustrates that the CSUE-SMI estimator tends toward a solution in which the structural shocks are less affected by a common volatility process.

\begin{figure}[ht]
\centering
\includegraphics[width=0.9\linewidth]{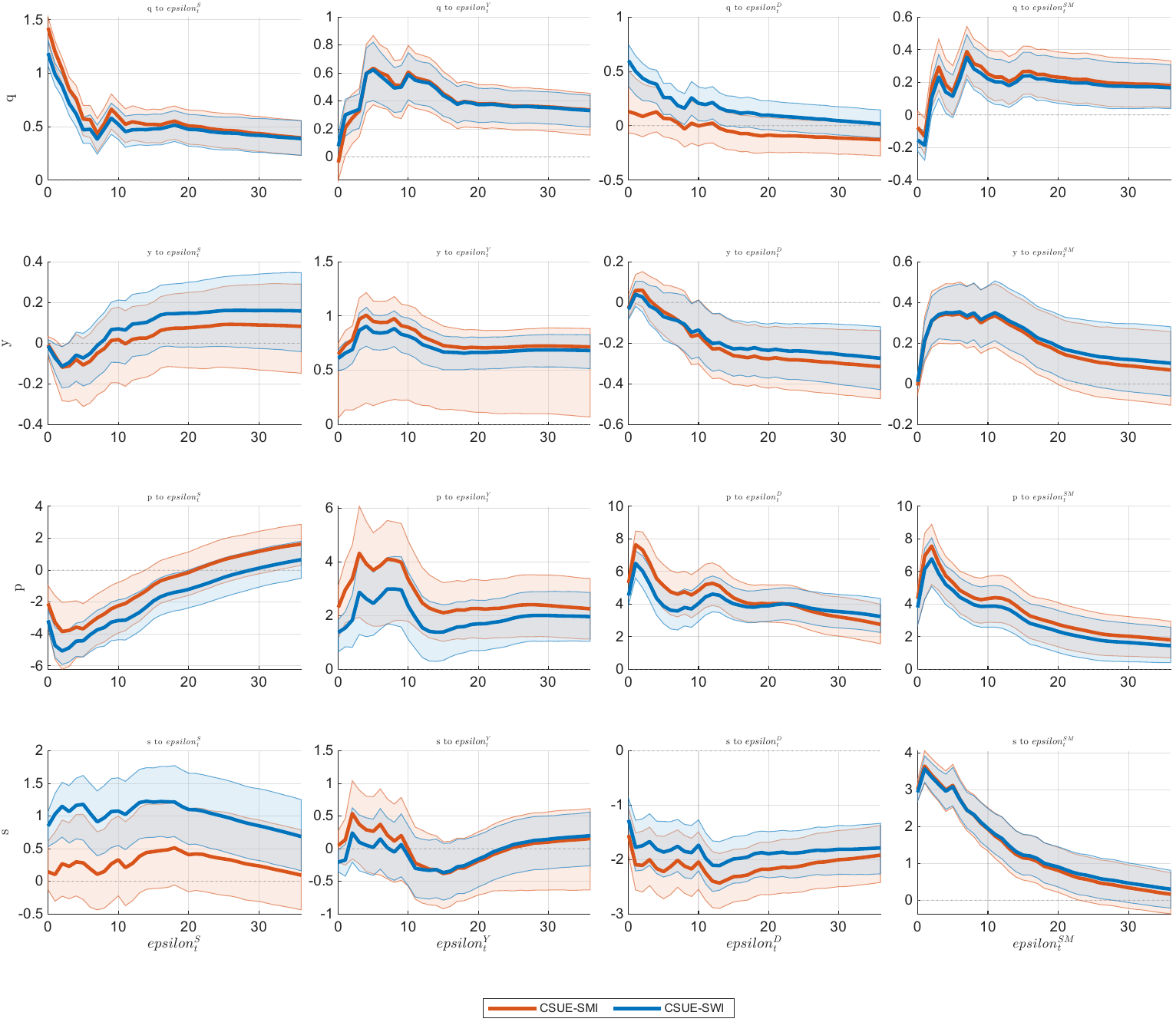}
\caption{Impulse responses from CSUE-SWI and CSUE-SMI. See Figure~\ref{fig:oil-irfs} for further details.}
\label{fig:fig_irf_comparison_2}
\end{figure}

Figure \ref{fig:fig_irf_comparison_2} illustrates the resulting consequences for the impulse response functions by comparing the responses obtained with the CSUE-SWI estimator in the main text with those obtained using the CSUE-SMI estimator. By avoiding solutions with large common volatility, the CSUE-SMI estimator produces two sizable differences in the IRFs. First, under CSUE-SMI, the impact stock-price response to a positive supply shock is smaller and becomes statistically insignificant. Second, impact oil production response to an oil-specific-demand shock is also smaller than under CSUE-SWI.

\end{document}